\documentclass[journal,twocolumn,10pt]{IEEEtran}
\usepackage{tikz}
\usepackage{array}
\usepackage{textcomp}
\usepackage{stfloats}
\usepackage{verbatim}
\usepackage{dirtytalk}
\usepackage{multirow}
\usepackage{rotating}
\usepackage{booktabs}
\usepackage{graphicx}
\usepackage{amsmath,amsfonts}
\usepackage[english]{babel}
\usepackage{amsthm}
\usepackage{bm}
\usepackage{hyperref}
\usepackage[none]{hyphenat}
\usepackage{url}
\usepackage{cite}
\usetikzlibrary{trees}
\usetikzlibrary{chains, fit}
\usepackage{listings}
\usepackage[subnum]{cases}

\usepackage{comment}
\usepackage{pgfplots}
\newcommand{\Herm}{{\mathrm{H}}}
\newcommand{\Trans}{{\mathrm{T}}}
\usepackage{pifont}
\newcommand{\sgn}{\mathrm{sgn}}
\newcommand{\VEC}{\mathrm{vec}}

\newtheorem{theorem}{Theorem}
\newtheorem{lemma}{Lemma}
\newtheorem{remark}{Remark}

\numberwithin{prop2}{section}

\newtheorem{corollary}{Corollary}
\usepackage{amssymb}

\newcommand{\Diag}[1]{\mathtt{diag}{{#1}}}
\newcommand{\BDiag}[1]{\mathtt{DIAG}{{#1}}}

\newcommand\copyrighttext{%
  \footnotesize \textcopyright 2025 IEEE. Personal use of this material is permitted. Permission from IEEE must be obtained for all other uses, in any current or future media, including
reprinting/republishing this material for advertising or promotional purposes, creating new collective works, for resale or redistribution to servers or lists, or
reuse of any copyrighted component of this work in other works.
}
\newcommand\copyrightnotice{%
\begin{tikzpicture}[remember picture,overlay]
\node[anchor=south,yshift=10pt] at (current page.south) {\fbox{\parbox{\dimexpr\textwidth-\fboxsep-\fboxrule\relax}{\copyrighttext}}};
\end{tikzpicture}%
}

\pgfplotsset{compat=1.18}
\begin{document}

\title{On Optimal MMSE Channel Estimation for One-Bit Quantized MIMO Systems}

\author{Minhua Ding, Italo Atzeni,~\IEEEmembership{Senior~Member,~IEEE,}\\ Antti T\"{o}lli,~\IEEEmembership{Senior Member,~IEEE,} and A. Lee Swindlehurst,~\IEEEmembership{Fellow,~IEEE}
\thanks{M. Ding, I. Atzeni, and A. T\"{o}lli are with the Centre for Wireless Communications, University of Oulu, Finland (e-mail: \{minhua.ding, italo.atzeni, antti.tolli\}@oulu.fi). A. L. Swindlehurst is with the Center for Pervasive Communications and Computing, University of California, Irvine, CA, USA (e-mail: swindle@uci.edu). This paper was presented in part at IEEE SPAWC 2024, Lucca, Italy, in September 2024~\cite{Ding_et_al_SPAWC2024}. This work was supported by the Research Council of Finland (336449 Profi6, 348396 HIGH-6G, 357504 EETCAMD, and 369116 6G~Flagship), by the European Commission (101095759 Hexa-X-II), and by the U. S. National Science Foundation (CCF-2225575).}}

\maketitle

\IEEEpeerreviewmaketitle

\copyrightnotice

\begin{abstract}
This paper focuses on the minimum mean squared error (MMSE) channel estimator for multiple-input multiple-output (MIMO) systems with one-bit quantization at the receiver side. Despite its optimality and significance in estimation theory, the MMSE estimator has not been fully investigated in this context due to its general nonlinearity and computational complexity. Instead, the typically suboptimal Bussgang linear MMSE (BLMMSE) channel estimator has been widely adopted. In this work, we develop a new framework to compute the MMSE channel estimator that hinges on the computation of the orthant probability of a multivariate normal distribution. Based on this framework, we determine a necessary and sufficient condition for the BLMMSE channel estimator to be optimal and thus equivalent to the MMSE estimator. Under the assumption of specific channel correlation or pilot symbols, we further utilize the framework to derive analytical expressions for the MMSE estimator that are particularly convenient for the computation when certain system dimensions become large, thereby enabling a comparison between the BLMMSE and MMSE channel estimators in these cases.
\end{abstract}


\section{Introduction}
Due to their ability to significantly enhance the spectral efficiency and reliability~\cite{Larsson_et_al, Rusek-et-al, Emil-etal-2017}, multiple-input multiple-output (MIMO) systems with a large number of antennas are employed in state-of-the-art wireless networks and are envisioned to assume an even more prominent role in future wireless generations~\cite{Raj20, Lozano_2021_WSA}. However, deploying such massive MIMO systems motivates the study of the use and impact of low-cost, energy-efficient hardware on the system performance. One of the main challenges in implementing a fully digital high-resolution massive MIMO system is the prohibitively high power consumption of the analog-to-digital converters (ADCs), which scales exponentially with the number of resolution bits~\cite{Walden_quant,Murmann_ADC_power}. Against this backdrop, there  has been intensive research on MIMO systems with low-resolution ADCs, e.g.,~\cite{Mezghani_Nossek_ISIT_2008, Jacobsson_TWC_UL, J_Mo_R_Heath,Bernando_2022, Choi_Love_2015, 1B_ML_data_detection, Shao_Ma_1B_detection_EM,Ivrlac_Nossek_2007, Choi_ML, Risi_Larsson_arxiv_paper, Jacobsson_et_al_2015_ICCW, Mesghani_Nossek_2010, Studer_Durisi_2016, Mo_asilomar_2014_CH, Wen_et_al, J_MO_CH_EST, Nguyen_2021, Nguyen_2023,Stein_unknown_threshold, Lee_et_al_Concatenated_MMSE,Y_Li_et_al_BLMMSE, Utschick_etal_2023,Atzeni_2022,Atz21a, Atz21b, Saxena_Swindlehurst}. In particular, one-bit quantization for MIMO systems has received special attention due to its energy efficiency and low complexity. An analysis of the capacity or capacity bounds of various one-bit quantized multi-antenna systems was carried out in~\cite{J_Mo_R_Heath} assuming perfect channel state information (CSI) at both the transmitter and receiver, whereas capacity bounds for one-bit MIMO systems with Gaussian channels and analog combining were studied in~\cite{Bernando_2022} under the same CSI assumption. Assuming perfect CSI at the receiver, maximum-likelihood (ML) data detection was studied for one-bit quantized MIMO systems in~\cite{Choi_Love_2015, 1B_ML_data_detection} and for one-bit MIMO-OFDM systems in~\cite{Shao_Ma_1B_detection_EM}.

Most massive MIMO systems require instantaneous CSI to facilitate the beamforming design. Channel estimation based on one-bit quantized measurements is thus of great importance~\cite{Ivrlac_Nossek_2007, Choi_ML, Risi_Larsson_arxiv_paper, Jacobsson_et_al_2015_ICCW, Mesghani_Nossek_2010, Studer_Durisi_2016, Wen_et_al, Nguyen_2021, Nguyen_2023, Stein_unknown_threshold,Y_Li_et_al_BLMMSE, Lee_et_al_Concatenated_MMSE,Utschick_etal_2023}. In this context, ML channel estimation was considered in~\cite{Ivrlac_Nossek_2007, Choi_ML}, whereas least squares channel estimation was studied in~\cite{Risi_Larsson_arxiv_paper, Jacobsson_et_al_2015_ICCW}. In~\cite{Mesghani_Nossek_2010, Studer_Durisi_2016}, channel estimation was studied using the maximum a posteriori approach~\cite{S-M-Kay_estimation}. Using a Bayesian framework, a joint channel-and-data estimation algorithm based on approximate message passing (AMP) for MIMO systems with low-resolution ADCs was proposed in~\cite{Wen_et_al}. Other methods based on AMP have also been investigated in~\cite{Mo_asilomar_2014_CH, J_MO_CH_EST}. More recently, channel estimation and data detection methods based on support vector machines and deep learning have been proposed in \cite{Nguyen_2021,Nguyen_2023} and references therein. Cram\'er-Rao bounds for channel estimation in one-bit quantized MIMO systems were studied in~\cite{Rao_et_al_CH_EST}. Channel estimation from one-bit quantized measurements with an unknown threshold was considered in~\cite{Stein_unknown_threshold}. The Bussgang linear minimum mean squared error (BLMMSE) channel estimator proposed in~\cite{Y_Li_et_al_BLMMSE} is based on the second-order statistics of the one-bit quantized observations as well as those of the fading channels~\cite{Bussgang_paper, Jacovitti_IT_complex_valued}. Its linear structure inspired subsequent signal processing methods applied to, e.g., data detection~\cite{Atzeni_2022,Atz21a} and precoding design~\cite{Saxena_Swindlehurst,Jacobsson_TWC_DL}. In addition, a generalized BLMMSE channel estimator was investigated in~\cite{Wan_generalized_BLM}.

In the study of channel estimation methods, various forms of the mean squared error (MSE) have often been adopted as a performance metric in numerical results~\cite{Ivrlac_Nossek_2007, Choi_ML, Mesghani_Nossek_2010, Lee_et_al_Concatenated_MMSE,Y_Li_et_al_BLMMSE, Utschick_etal_2023, Wen_et_al, Nguyen_2021, Nguyen_2023,Stein_unknown_threshold}. Despite its optimality in terms of MSE and its significance in estimation theory, the minimum MSE (MMSE) channel estimator, also known as the conditional mean estimator~\cite{S-M-Kay_estimation, Curry_R_Est_Control_quant, Utschick_etal_2023}, has not been fully investigated for one-bit quantized MIMO systems. The main challenge in computing the general MMSE channel estimator lies in dealing with the seemingly intractable multi-dimensional integration involving the multivariate normal (MVN) distribution~\cite{S-M-Kay_estimation, Curry_R_Est_Control_quant}. In~\cite{Wen_et_al},   while the main focus was on AMP-based joint channel-and-data estimation for MIMO systems with low-resolution ADCs, a closed-form expression of the MMSE channel estimator was derived for low-resolution quantized single-input single-output (SISO) systems with a single pilot symbol, which matches the results in~\cite{Ziniel_2015} with one-bit quantization. The result in~\cite{Wen_et_al} was used in the study of a concatenated channel estimation method  in~\cite{Lee_et_al_Concatenated_MMSE}. However, for one-bit quantized systems, the channel estimator in~\cite{Lee_et_al_Concatenated_MMSE} reduces to the BLMMSE estimator. In~\cite{Utschick_etal_2023}, assuming a spatially white single-input multiple-output (SIMO) channel, the equivalence between the BLMMSE and the optimal MMSE channel estimators with one-bit ADCs was established for noisy SIMO systems with a single pilot symbol as well as for noiseless SIMO systems with a sequence of MSE-optimal pilot symbols. However, a general condition for the equivalence of the two estimators remains unknown and the explicit computation of the MMSE channel estimate still appears elusive.

In this work, we provide further results on the optimal MMSE channel estimator with one-bit ADCs for Rayleigh fading MIMO channels with additive white Gaussian noise (AWGN). We consider a point-to-point MIMO system to focus on the fundamental estimation problem, although our results can be framed for a multi-user uplink system with single-antenna users and a multi-antenna base station. Despite the inherent complexity of the estimation problem, we obtain several new results, which are summarized as follows.
\begin{itemize}
\item We first develop a general framework 
for  explicitly computing the MMSE channel estimate, which is valid for arbitrary channel correlation and arbitrary pilot matrices. The computation of the MMSE estimate in our framework is connected to the computation of the orthant probability of a MVN distribution. The Gaussian orthant probability has been extensively investigated in the literature, e.g., in~\cite{Abrahamsom_64, Childs-1967, Plackett_54, Sun_1988, Tong_MVN_book, Genz_algorithm, Takemura}, and has been used in various engineering fields, e.g., in the time series analysis of binary data~\cite{Keenan_1982}, in the study of zero crossing properties of speech signals~\cite{Sinn_Keller_cov_of_zero_crossing}, and in one-bit direction-of-arrival estimation~\cite{Stein_WSA_2016}. Thus, our framework connects the investigation at hand to a wider class of problems that involve the Gaussian orthant probability. Through the Gaussian orthant probability, our expression for the MMSE channel estimator reveals the essential challenge in its computation and sheds light on the interplay between the channel correlation and the pilot design on the complexity.
\item We determine a necessary and sufficient condition for the equivalence of the BLMMSE and MMSE channel estimates in one-bit quantized MIMO systems. Moreover, we examine specific cases where this equivalence holds. The simplest case where the BLMMSE channel estimate becomes suboptimal is identified to correspond to a SIMO system with a single pilot symbol and full real-valued channel correlation among three receive antennas. For this specific case, we present a closed-form expression for the optimal nonlinear MMSE channel estimate.
\item We derive an expression for the MMSE channel estimate for SIMO channels with a single pilot symbol assuming equal real-valued correlation among all the receive antennas. In addition, we derive an expression for the optimal MMSE channel estimate for SISO channels or spatially white SIMO channels with multiple real-valued pilot symbols. These expressions require only integration  over a single variable and can thus be efficiently computed when the number of receive antennas or pilot symbols becomes large.
\end{itemize}

A preliminary version of this work was presented in~\cite{Ding_et_al_SPAWC2024}, where we discussed the optimality of the BLMMSE channel estimator and provided, without proofs, a version of Theorems~\ref{theo: Theorem_BLMMSE_opt} and~\ref{theo: SIMO_N3_nonL}. In this paper, we complement the contribution of~\cite{Ding_et_al_SPAWC2024} with streamlined details and proofs as well as additional specific cases. In addition, Theorems~\ref{theo: SIMO_general_N_special_corr} and~\ref{theo: theo_rank_one_SISO} are entirely new.

\emph{Outline.} The rest of the paper is organized as follows. Section~\ref{sec: sys_model_prob_state} describes the system model and the goals of this investigation. Section~\ref{sec: framework} first provides a background on the Gaussian orthant probability and then develops the framework for computing the MMSE channel estimate in one-bit quantized MIMO systems. The condition for the optimality of the BLMMSE channel estimate is presented in Section~\ref{sec: BLMMSE_opt_results}. Various results on the nonlinear MMSE channel estimate are discussed in Section~\ref{sec: non-L results}, whereas numerical results are provided in Section~\ref{sec: simulation results}. Finally, Section~\ref{sec: conclusions} concludes the paper. Detailed derivations and proofs are given in the respective appendices.

\emph{Notation.} Lowercase letters, bold lowercase letters, and bold uppercase letters represent scalars, vectors, and matrices, respectively. The $k$-th element of the vector $\mathbf{x}$ and the $(i, k)$-th element of the matrix $\mathbf{X}$ are denoted by $x_k$ and $[\mathbf{X}]_{ik}$, respectively. The operators $(\cdot)^*, (\cdot)^\Trans$, and $(\cdot)^\Herm$ denote complex conjugate, transpose, and Hermitian transpose, respectively. The $n\times n$ identity matrix is written as $\mathbf{I}_{n}$. The operator $\otimes$ represents the Kronecker product, $|\mathbf{X}|$ denotes the determinant of the square matrix $\mathbf{X}$, $\VEC(\cdot)$ denotes vectorization, $\mathrm{tr}(\cdot)$ represents the trace, and $\|\mathbf{x}\|=\sqrt{\mathbf{x}^\Herm\mathbf{x}}$. The sign function is indicated by $\sgn (\cdot)$. $\Re(\cdot)$ and $\Im(\cdot)$ represent the real and imaginary parts, respectively, $|x|$ denotes the modulus of $x$, and $j=\sqrt{-1}$. The sets of real-valued non-negative, real-valued, and complex-valued $n$-dimensional vectors are denoted by $\mathbb{R}_{+}^n$, $\mathbb{R}^n$, and $\mathbb{C}^n$, respectively, whereas the sets of real- and complex-valued $m\times n$ matrices are denoted by $\mathbb{R}^{m\times n}$ and $\mathbb{C}^{m\times n}$, respectively. The real- and complex-valued MVN distributions with mean vector $\boldsymbol{\mu}$ and covariance matrix $\boldsymbol{\Sigma}$ are indicated by $\mathcal{N}(\boldsymbol{\mu}, \boldsymbol{\Sigma})$ and $\mathcal{CN}(\boldsymbol{\mu}, \boldsymbol{\Sigma})$, respectively. The operators $\mathsf{E}(\cdot)$ and $\Pr(\cdot)$ represent statistical expectation and probability, respectively. In addition, $[z_k]_{k=1}^m$ denotes the vector $[z_1 \ldots z_m]^{\Trans}$, $\Diag(\mathbf{z})$ is a diagonal matrix with the elements of $\mathbf{z}$ on its diagonal, $\widetilde{\Diag{}}(\mathbf{X})$ is a diagonal matrix with the diagonal elements of the square matrix $\mathbf{X}$ on its diagonal, and $\BDiag(\mathbf{Y}_1, \ldots, \mathbf{Y}_n)$ is a block-diagonal matrix with blocks $\mathbf{Y}_1, \ldots, \mathbf{Y}_n$. Lastly, $\mathbf{X}_{-k}$ is the $(n-1)\times (n-1)$ matrix obtained by deleting the $k$-th row and column of the $n\times n$ matrix $\mathbf{X}$, whereas $\mathbf{x}_{-k}$ is the $(n-1)\times 1$ vector obtained by deleting the $k$-th element of the $n\times 1$ vector $\mathbf{x}$.

\section{System Model and Problem Statement}\label{sec: sys_model_prob_state}

\subsection{System Model}\label{sec: system_model}

Consider a general MIMO system with $N_T$ transmit antennas and $N_R$ receive antennas. Assume slow flat Rayleigh fading and let $\mathbf{H}\in\mathbb{C}^{N_R\times N_T}$ denote the channel between the transmitter and the receiver. 

For the channel estimation, $\tau$ pilot symbols per antenna are transmitted, which are collectively denoted by $\mathbf{S}^\Trans=[\mathbf{s}_1 \ldots \mathbf{s}_\tau]\in\mathbb{C}^{N_T\times\tau}$. We define the received signal at the input of the ADCs as
\begin{align}
\mathbf{B} = \mathbf{H}\mathbf{S}^\Trans+\mathbf{N} =[\mathbf{b}_1 \ldots \mathbf{b}_\tau] \in\mathbb{C}^{N_R\times\tau}, \label{sm: mat_SM}
\end{align}
where $\mathbf{N}=[\mathbf{n}_1 \ldots \mathbf{n}_\tau] \in\mathbb{C}^{N_R\times\tau}$ represents the noise. Furthermore, we introduce
\begin{align}
\mathbf{b} = \VEC(\mathbf{B}) =(\mathbf{S}\otimes\mathbf{I}_{N_R})\mathbf{h} + \mathbf{n}=\mathbf{A}\mathbf{h}+\mathbf{n} \in\mathbb{C}^{\tau N_R}, \label{sm: vec_SM}
\end{align}
with $\mathbf{h}=\VEC(\mathbf{H}) \in\mathbb{C}^{N_TN_R}$, $\mathbf{n}=\VEC(\mathbf{N}) \in\mathbb{C}^{\tau N_R}$, and 
\begin{align}
\mathbf{A}=\mathbf{S}\otimes\mathbf{I}_{N_R} \in\mathbb{C}^{(\tau N_R)\times (N_TN_R)}.\label{eqn: matrix_A}
\end{align}
Based on the channel assumption, we have $\mathbf{h}\sim\mathcal{CN}(\mathbf{0}, \boldsymbol{\Sigma})$, where $\boldsymbol{\Sigma}$ represents the $(N_TN_R)\times (N_TN_R)$ channel covariance matrix. Hence, the probability density function of the channel is given by
\begin{align}
p(\mathbf{h})=\frac{1}{\pi^{N_T N_R}|\boldsymbol{\Sigma}|}\exp\left\{-\mathbf{h}^\Herm\boldsymbol{\Sigma}^{-1}\mathbf{h}\right\}. \label{eqn: prob_density_h}
\end{align}
We assume additive, spatially and temporally white, and circularly symmetric complex Gaussian noise, i.e.,  $\mathbf{n}\sim\mathcal{CN}(\mathbf{0}, \sigma^2\mathbf{I}_{\tau N_R})$. We denote the element-wise memoryless one-bit quantization operation as
\begin{align}
\mathcal{Q}_{\rm 1bit}(\cdot) = \sgn\big(\Re(\cdot)\big)+j\sgn\big(\Im(\cdot)\big).
\label{eqn: sm_QUANT}
\end{align}
Then, the quantized received signal (at the output of the ADCs) is given by
\begin{align}
\mathbf{r} =\mathcal{Q}_{\rm 1bit}(\mathbf{b})\in\mathbb{C}^{\tau N_R}.
\label{eqn: quantized_y}
\end{align} 

Based on the quantized output $\mathbf{r}$, the MMSE estimate of $\mathbf{h}$ is given by $\hat{\mathbf{h}}_{\rm MMSE} =\mathsf{E}(\mathbf{h}|\mathbf{r})=\int_{\mathbb{C}^{N_T N_R}}\mathbf{h}p(\mathbf{h}|\mathbf{r}){\rm d}\mathbf{h}$, with $p(\mathbf{h}|\mathbf{r})=\Pr(\mathbf{r}|\mathbf{h})p(\mathbf{h})/\Pr(\mathbf{r})$~\cite{S-M-Kay_estimation}. Therefore,
\begin{align}
\hat{\mathbf{h}}_{\rm MMSE} =\frac{1}{\Pr(\mathbf{r})}\displaystyle\int_{\mathbb{C}^{N_T N_R}}\mathbf{h}\Pr(\mathbf{r}|\mathbf{h})p(\mathbf{h}){\rm d}\mathbf{h}, \label{eqn: h_MMSE_txt_formula}
\end{align}
with
\begin{align}
\Pr(\mathbf{r})= \int_{\mathbb{C}^{N_TN_R}}\Pr(\mathbf{r}|\mathbf{h}) p(\mathbf{h}) {\rm d}\mathbf{h}.\label{eqn: prob_r_first_appear}
\end{align}

\subsection{Problem Statement}
We now state the goals of our investigation.
\begin{itemize}
\item  To the best of our knowledge, there are no explicit details in the literature on how to compute \eqref{eqn: h_MMSE_txt_formula} in the context of MIMO systems with one-bit ADCs. Therefore, we first develop a novel framework for its computation. Then, we determine a necessary and sufficient condition under which the MMSE channel estimate is linear, i.e., a necessary and sufficient condition for the BLMMSE channel estimator to be MSE-optimal.
\item We further identify the simplest case where the 
nonlinear MMSE channel estimate admits a closed-form expression. Despite the generic complexity of computing \eqref{eqn: h_MMSE_txt_formula}~\cite{Wen_et_al,Utschick_etal_2023}, under specific assumptions on the channel correlation or pilot symbols, we aim at obtaining computationally efficient analytical expressions for $\hat{\mathbf{h}}_{\rm MMSE}$ that facilitate insight into its performance when certain system dimensions, i.e., the number of receive antennas or pilot symbols, become large. 
\end{itemize}

\section{A Framework for the MMSE Channel Estimation of One-Bit Quantized MIMO Systems}\label{sec: framework}

In this section, we examine the structure of the MMSE channel estimator with one-bit quantized observations and reformulate the corresponding conditional mean expression in a form that involves the Gaussian orthant probability~\cite{Abrahamsom_64, Keenan_1982, Sinn_Keller_cov_of_zero_crossing}. This new expression will then pave the way for deriving the results in Sections~\ref{sec: BLMMSE_opt_results} and \ref{sec: non-L results}. 

\subsection{Preliminaries on the Orthant Probability of a Real-Valued MVN Distribution} \label{Sec: OrthantP_basics}

To simplify the subsequent derivations, we first introduce the orthant probability of the real-valued MVN distribution, which is defined as~\cite{Abrahamsom_64, Keenan_1982, Sinn_Keller_cov_of_zero_crossing}
\begin{align}
\mathcal{P}(\boldsymbol{\Psi})= \frac{1}{(2\pi)^{\frac{L}{2}}\left|\boldsymbol{\Psi}\right|^{\frac{1}{2}}}\displaystyle \int_{\mathbb{R}_+^{L}} 
\exp\left\{-\frac{1}{2}\mathbf{u}^\Trans \boldsymbol{\Psi}^{-1}\mathbf{u}\right\} {\rm d}\mathbf{u}, \label{eqn: orthant_Prob_formula}
\end{align}
where $\boldsymbol{\Psi}$ denotes the $L\times L$ covariance matrix in $\mathcal{N}(\mathbf{0}, \boldsymbol{\Psi})$. Let us define the standardized version of $\boldsymbol{\Psi}$ as
\begin{align}
\boldsymbol{\Psi}_{\rm std}= \big(\widetilde{\Diag{}}(\boldsymbol{\Psi})\big)^{-\frac{1}{2}}\boldsymbol{\Psi}\big(\widetilde{\Diag{}}(\boldsymbol{\Psi})\big)^{-\frac{1}{2}},\label{eqn: standardized_cov}
\end{align}
which denotes the covariance matrix of the standardized version of $L$  random variables whose original covariance matrix is $\boldsymbol{\Psi}$ in \eqref{eqn: orthant_Prob_formula}~\cite{Johnson_Wichern_stats}. Clearly, the diagonal elements of $\boldsymbol{\Psi}_{\rm std}$ are all equal to one and the off-diagonal elements represent the correlation coefficients between the random variables. 

Through a change of variables, one can show that~\cite{Childs-1967}
\begin{align}
\mathcal{P}(\boldsymbol{\Psi})= \mathcal{P}(\boldsymbol{\Psi}_{\rm std}). \label{eqn: OrthantP_identity}
\end{align} 
Based on \eqref{eqn: OrthantP_identity}, it is  straightforward to see that
 \begin{align}
\mathcal{P}(\boldsymbol{\Psi}) &=\mathcal{P}(c_{0}\boldsymbol{\Psi})  \label{eqn: OrthantP_scaling}
\end{align}
for a constant $c_0>0$ and that, for a diagonal $\boldsymbol{\Psi}$, we have
\begin{align}
\mathcal{P}(\boldsymbol{\Psi})=2^{-L}.  
\label{eqn: orthantP_diag_mat}
\end{align}
For $i, k=1, \ldots, L$ and $i\neq k$, let 
\begin{align}\label{eqn: orthantP_corr_coeff}
\left[\boldsymbol{\Psi}_{\rm std}\right]_{ik} =\psi_{ik} \in(-1, 1). 
\end{align}
It was shown in~\cite{Abrahamsom_64, Childs-1967, Tong_MVN_book} that, for $L=2$ and $L=3$, we have
\begin{align}
\mathcal{P}(\boldsymbol{\Psi}) & = 
\frac{1}{4}+\frac{\arcsin(\psi_{12})}{2\pi}, \label{eqn: OrthantP_N2} \\
\mathcal{P}(\boldsymbol{\Psi}) & = 
\frac{1}{8}+\frac{\sum_{i=1}^2\sum_{k=i+1}^3\arcsin(\psi_{ik})}{4\pi},
\label{eqn: OrthantP_N3}
\end{align}
respectively. Furthermore, for $L=4$, an explicit expression for $\mathcal{P}(\boldsymbol{\Psi})$ based on~\cite{Childs-1967} is provided in \eqref{eqn: OrthantP4} (see Appendix~\ref{App: OrthantP_basics_P4}), which requires finite integration with one variable. There is no general closed-form expression for $\mathcal{P}(\boldsymbol{\Psi})$ when $L\geq 4$~\cite{Childs-1967, Sun_1988, Plackett_54, Takemura}.

In~\cite[Eq. (6)]{Childs-1967}, an expression for the orthant probability was derived based on characteristic functions~\cite{Papoullis}, where the number of terms to be evaluated grows exponentially with $L$ and the complexity of evaluating some individual integral terms also grows with $L$. For $L>4$, one can use Monte Carlo methods to compute \eqref{eqn: orthant_Prob_formula}, e.g., the algorithm in~\cite{Genz_algorithm} that was used in~\cite{Utschick_etal_2023}. However, Monte Carlo methods also incur a high computational load when $L$ increases and the computation required for \eqref{eqn: orthant_Prob_formula}  still seems intractable when $L$ is large~\cite{Takemura, Miwa_et_al}.

\subsection{Framework for Computing the MMSE Channel Estimate for One-Bit Quantized MIMO Systems} \label{subsection: Expression for the MMSE channel estimate}

Recall \eqref{sm: vec_SM} and \eqref{eqn: sm_QUANT}--\eqref{eqn: quantized_y}, and let $\mathbf{r}=\mathbf{r}_{\mathsf{R}}+j\mathbf{r}_{\mathsf{I}}$, with 
\begin{align}
\mathbf{r}_{\mathsf{R}}=\Re(\mathbf{r})=\sgn\big(\Re(\mathbf{b})\big), \quad \mathbf{r}_{\mathsf{I}}=\Im(\mathbf{r})=\sgn\big(\Im(\mathbf{b})\big). \label{eqn:quantized_output_vector}
\end{align}
Then, $r_{\mathsf{R}, k}=\Re(r_k)$ and $r_{\mathsf{I}, k}=\Im(r_k)$, with $r_{\mathsf{R}, k}, r_{\mathsf{I}, k}\in\{\pm 1\}$, for $k=1, \ldots, \tau N_R$. Correspondingly, we have
\begin{align}
r_k & = r_{\mathsf{R}, k} + j r_{\mathsf{I}, k} = \sgn \big(\Re(b_k)\big)+j\sgn\big(\Im(b_k)\big) \nonumber\\
& = \sgn\big(\Re(\mathbf{a}_k^\Trans\mathbf{h})+\Re(n_k)\big) + j\sgn\big(\Im(\mathbf{a}_k^\Trans\mathbf{h})+\Im(n_k)\big), \label{eqn: k_th_scalar_r_k}
\end{align} 
where $\mathbf{a}_k^\Trans$ denotes the $k$-th row of $\mathbf{A}$.
Let
\begin{align}
\boldsymbol{\Lambda}_{\mathsf{R}} = \Diag(\mathbf{r}_{\mathsf{R}}), \quad \boldsymbol{\Lambda}_{\mathsf{I}} = \Diag(\mathbf{r}_{\mathsf{I}}). \label{eqn:quantized_output_diag_mat}
\end{align}
Note that 
$\boldsymbol{\Lambda}_{\mathsf{R}}\boldsymbol{\Lambda}_{\mathsf{R}}=\boldsymbol{\Lambda}_{\mathsf{I}}\boldsymbol{\Lambda}_{\mathsf{I}}=\mathbf{I}_{\tau N_R}$. Denote the covariance matrix of $\mathbf{b}$ in \eqref{sm: vec_SM} as
\begin{align}
\boldsymbol{\Omega}_{\rm b}= \mathbf{A}\boldsymbol{\Sigma}\mathbf{A}^\Herm+\sigma^2\mathbf{I}_{\tau N_R}\in\mathbb{C}^{(\tau N_R)\times(\tau N_R)}
\label{eqn: corr_mat_of_b}
\end{align}
and further define
\begin{align}
\mathbf{D}_{\mathsf{R}}  &= \Re(\boldsymbol{\Omega}_{\rm b}^{-1}) = \mathbf{D}_{\mathsf{R}}^\Trans \in\mathbb{R}^{(\tau N_R)\times(\tau N_R)}, \label{eqn: matrix_DR}
\\
\mathbf{D}_{\mathsf{I}}  &=  \Im(\boldsymbol{\Omega}_{\rm b}^{-1}) = -\mathbf{D}_{\mathsf{I}}^\Trans \in\mathbb{R}^{(\tau N_R)\times(\tau N_R)}, \label{eqn: matrix_DI}
\\
\mathbf{C}  &= \begin{bmatrix}
\boldsymbol{\Lambda}_{\mathsf{R}}&\mathbf{0} \\
\mathbf{0}& \boldsymbol{\Lambda}_{\mathsf{I}}
\end{bmatrix}\begin{bmatrix}
\mathbf{D}_{\mathsf{R}} & \mathbf{D}_{\mathsf{I}}^\Trans 
\\
\mathbf{D}_{\mathsf{I}} & \mathbf{D}_{\mathsf{R}}
\end{bmatrix}\begin{bmatrix}
\boldsymbol{\Lambda}_{\mathsf{R}}&\mathbf{0} \\
\mathbf{0}& \boldsymbol{\Lambda}_{\mathsf{I}}
\end{bmatrix}\nonumber
\\&=\begin{bmatrix}
\boldsymbol{\Lambda}_{\mathsf{R}}\mathbf{D}_{\mathsf{R}}\boldsymbol{\Lambda}_{\mathsf{R}} & \boldsymbol{\Lambda}_{\mathsf{R}}\mathbf{D}_{\mathsf{I}}^\Trans\boldsymbol{\Lambda}_{\mathsf{I}}\\
\boldsymbol{\Lambda}_{\mathsf{I}}\mathbf{D}_{\mathsf{I}}\boldsymbol{\Lambda}_{\mathsf{R}} & \boldsymbol{\Lambda}_{\mathsf{I}}\mathbf{D}_{\mathsf{R}}\boldsymbol{\Lambda}_{\mathsf{I}}
\end{bmatrix} \in\mathbb{R}^{(2\tau N_R)\times (2\tau N_R)}. \label{eqn: Matrix_C} 
\end{align}
The MMSE channel estimate can be expressed as
\begin{align}
\hat{\mathbf{h}}_{\rm MMSE} & = \boldsymbol{\Sigma}\mathbf{A}^\Herm\boldsymbol{\Omega}_{\rm b}^{-1} \left[\boldsymbol{\Lambda}_{\mathsf{R}} \ j\boldsymbol{\Lambda}_{\mathsf{I}}\right] \frac{\displaystyle \int_{\mathbb{R}_+^{2\tau N_R}} \mathbf{z} \exp\{-\mathbf{z}^\Trans\mathbf{C}\mathbf{z}\} {\rm d}\mathbf{z}}{\displaystyle\int_{\mathbb{R}_+^{2\tau N_R}} \exp\{-{\mathbf{z}}^\Trans \mathbf{C}{\mathbf{z}}\} {\rm d}{\mathbf{z}}} \label{eqn: compact_MMSE_form1}
\end{align}
(see derivations in Appendix~\ref{App: main_framework_derivation}). Note that the quotient of integrals in \eqref{eqn: compact_MMSE_form1} denotes the mean of a truncated MVN distribution $\mathcal{N}(\mathbf{0}, (2\mathbf{C})^{-1})$ with $\mathbb{R}_{+}^{2\tau N_R}$ as its support. 

Next, we express the integrals in \eqref{eqn: compact_MMSE_form1} using the orthant probability defined in \eqref{eqn: orthant_Prob_formula}. Through the details given in Appendix~\ref{App: deriving_h_MMSE_using_Orthant_P}, we have
 \begin{align}
\hat{\mathbf{h}}_{\rm MMSE} = \frac{\boldsymbol{\Sigma}\mathbf{A}^\Herm\boldsymbol{\Omega}_{\rm b}^{-1} \left[\boldsymbol{\Lambda}_{\mathsf{R}} \ j\boldsymbol{\Lambda}_{\mathsf{I}}\right] \mathbf{C}^{-1}\big(\widetilde{\Diag{}}(\mathbf{C}^{-1})\big)^{-\frac{1}{2}}\mathbf{g}}
{2\sqrt{\pi}\mathcal{P}\left(\mathbf{C}^{-1}\right)}, \label{eqn: reduction_2}
\end{align}
where the $k$-th element of $\mathbf{g}$ is given by
\begin{align}
g_k =\mathcal{P}\left((\mathbf{C}_{-k})^{-1}\right), \label{eqn: new_g_k}
\end{align}
and $\mathbf{C}_{-k}\in\mathbb{R}^{(2\tau N_R-1)\times (2\tau N_R-1)}$ is the matrix obtained by deleting the $k$-th row and column of $\mathbf{C}$, for $k=1, \ldots, 2\tau N_R$.

 \vspace{2mm}
 \begin{remark} 
 Note that \eqref{eqn: reduction_2}--\eqref{eqn: new_g_k} are valid for general noisy Rayleigh fading MIMO channels with arbitrary channel correlation and pilot matrices. In particular, the results here subsume~\cite[Theorem 1]{Utschick_etal_2023} as a special case, which considers a real-valued noiseless SIMO channel with $\tau=N_T=1$, $s=1$, $\mathbf{A}=\mathbf{I}_{N_R}$, $\boldsymbol{\Omega}_{\rm b}=\boldsymbol{\Sigma}$, $\boldsymbol{\Lambda}_{\mathsf{I}}=\mathbf{0}$, and $\mathbf{C}=\boldsymbol{\Lambda}_{\mathsf{R}}\boldsymbol{\Sigma}^{-1}\boldsymbol{\Lambda}_{\mathsf{R}}$. In the expression for $\hat{\mathbf{h}}_{\rm MMSE}$ used in~\cite{Utschick_etal_2023}, the  regions for multi-dimensional integration are defined only implicitly. In contrast, the integration required in \eqref{eqn: compact_MMSE_form1}--\eqref{eqn: new_g_k} is clearly specified. The quantized output in \eqref{eqn:quantized_output_vector} or \eqref{eqn:quantized_output_diag_mat} appears both outside the integrals and in the integrands through $\mathbf{C}$ and $\mathbf{C}_{-k}$, for $k=1, \ldots, 2\tau N_R$.
\end{remark}

\begin{remark}\label{remark: on pairwise_quantized_op_in_C}
The diagonal elements of $\mathbf{C}$ are not affected by the quantized output. Each off-diagonal element of $\mathbf{C}$ is affected by exactly two different elements from $[\mathbf{r}_{\mathsf{R}}^\Trans \ \mathbf{r}_{\mathsf{I}}^\Trans]^\Trans$. The same comments also apply to $\mathbf{C}_{-k}$, for $k=1, \ldots, 2\tau N_R$. These observations are instrumental for deriving simplified expressions for $\hat{\mathbf{h}}_{\rm MMSE}$ under specific assumptions on the channel correlation or pilot symbols (see Theorems~\ref{theo: SIMO_general_N_special_corr} and~\ref{theo: theo_rank_one_SISO} in Section~\ref{sec: non-L results}). 
\end{remark}

\begin{remark}\label{Remark: effect of correlation and pilots}
From \eqref{eqn: reduction_2}--\eqref{eqn: new_g_k} and the discussions in Section~\ref{Sec: OrthantP_basics}, one can immediately see the inherent complexity of computing the general $\hat{\mathbf{h}}_{\rm MMSE}$ when the system dimensions, i.e., $\tau$ and $N_R$, become large. The number of transmit antennas $N_T$ is not involved in computing the orthant probabilities. Furthermore, from \eqref{eqn: matrix_A}, \eqref{eqn: Matrix_C}--\eqref{eqn: compact_MMSE_form1}, and \eqref{eqn: reduction_2}--\eqref{eqn: new_g_k}, the channel covariance matrix $\boldsymbol{\Sigma}$ and the pilot matrix $\mathbf{S}^\Trans$ jointly determine the computational complexity of $\hat{\mathbf{h}}_{\rm MMSE}$ through  $\boldsymbol{\Omega}_{\rm b}$, which is embedded in $\mathbf{C}$ in \eqref{eqn: Matrix_C}. 
\end{remark}

\section{Condition for the Optimality of the BLMMSE Channel Estimator}\label{sec: BLMMSE_opt_results}
 In this section, we establish a general condition for  the BLMMSE channel estimator to be MSE-optimal and thus equivalent to the optimal MMSE estimator in one-bit quantized Rayleigh fading  MIMO systems with arbitrary channel correlation and pilot matrices. 

For the system model in Section~\ref{sec: system_model}, the BLMMSE channel estimator proposed in~\cite{Y_Li_et_al_BLMMSE} can be expressed as
\begin{align}
\hat{\mathbf{h}}_{\rm BLM} & = \frac{\sqrt{\pi}}{2}\boldsymbol{\Sigma}\mathbf{A}^\Herm\mathbf{D}_\Omega^{-\frac{1}{2}}\mathbf{T}_{\rm BLM}^{-1}\mathbf{r},
\label{eqn: full_BLMMSE_formula}
\end{align}
with
\begin{align}
\mathbf{D}_{\Omega} &=\widetilde{\Diag{}}(\boldsymbol{\Omega}_{\rm b}), \label{eqn: D_omega}
\\
    \mathbf{T}_{\rm BLM} &= \arcsin\big(\mathbf{D}_\Omega^{-\frac{1}{2}}\Re\left(\boldsymbol{\Omega}_{\rm b}\right)\mathbf{D}_\Omega^{-\frac{1}{2}}
\big) \nonumber \\
& \phantom{=} \ + j\arcsin\big(\mathbf{D}_\Omega^{-\frac{1}{2}}\Im\left(\boldsymbol{\Omega}_{\rm b}\right)\mathbf{D}_\Omega^{-\frac{1}{2}}
\big).
\end{align}
The resulting per-antenna MSE obtained by using $\hat{\mathbf{h}}_{\rm BLM}$ is given by~\cite{Y_Li_et_al_BLMMSE, Atzeni_2022}
\begin{align}
    \textrm{MSE}^{\rm BLM} &=\frac{1}{N_TN_R} \mathsf{E}\big(\|\mathbf{h}-\hat{\mathbf{h}}_{\rm BLM}\|^2\big) \nonumber
    \\&=\frac{\mathrm{tr}(\boldsymbol{\Sigma}) - \mathrm{tr}(\boldsymbol{\Sigma}\mathbf{A}^\Herm\mathbf{D}_\Omega^{-\frac{1}{2}}\mathbf{T}_{\rm BLM}^{-1}\mathbf{D}_\Omega^{-\frac{1}{2}}\mathbf{A}\boldsymbol{\Sigma})}{N_TN_R}. \label{eqn: MSE_per_antenna_general}
\end{align}

Before presenting the condition for the BLMMSE channel estimate to be MSE-optimal in Theorem~\ref{theo: Theorem_BLMMSE_opt} below, we introduce two lemmas that facilitate  its proof.

\begin{lemma}
\label{lemma: OrthantP_2_linearity}
Let $\mathbf{q}=[q_1 \; q_2]^\Trans$, with $q_1, q_2\in\{\pm 1\}$, and let $\boldsymbol{\Lambda}_\mathbf{q} = \Diag(\mathbf{q})$. Consider the $2\times 2$ covariance-type matrix 
\begin{align}
    \boldsymbol{\Phi}=\boldsymbol{\Lambda}_\mathbf{q}
    \begin{bmatrix}
        \breve{\sigma}_{1}^2  & \phi_{12}\breve{\sigma}_{1}\breve{\sigma}_{2}\\ \phi_{12}\breve{\sigma}_{1}\breve{\sigma}_{2} & \breve{\sigma}_{2}^2
    \end{bmatrix}
    \boldsymbol{\Lambda}_\mathbf{q}, \label{eqn: Matrix_Phi_in_lemma1}
\end{align}
with $\phi_{12}\in(-1, 1)$. Then, the vector 
\begin{align}
\boldsymbol{\Lambda}_\mathbf{q} \frac{\displaystyle\int_{\mathbb{R}_+^2}\mathbf{u}\exp\left\{-\mathbf{u}^\Trans\boldsymbol{\Phi}\mathbf{u}\right\}{\rm d}\mathbf{u}}{\displaystyle\int_{\mathbb{R}_+^2}\exp\left\{-\mathbf{u}^\Trans\boldsymbol{\Phi}\mathbf{u}\right\}{\rm d}\mathbf{u}}
\label{eqn: key_expression_in_lemma1_for_theorem}
\end{align}
is linear in $\mathbf{q}$.
\end{lemma}

\begin{IEEEproof}
    See Appendix~\ref{App_proof_Lemma}.
\end{IEEEproof}

\smallskip

With a slight abuse of notation, we further introduce the following lemma.

\begin{lemma}\label{lemma: OrthantP_3_nonL}
Let $\mathbf{q}=[q_1 \; q_2 \; q_3]^\Trans$, with $q_1, q_2, q_3\in\{\pm 1\}$, and let $\boldsymbol{\Lambda}_\mathbf{q} = \Diag(\mathbf{q})$. Consider the $3\times 3$ covariance-type matrix $\boldsymbol{\Phi}$, with $[\boldsymbol{\Phi}]_{ii}=\breve{\sigma}_{i}^2$, for $i=1, 2, 3$, $[\boldsymbol{\Phi}]_{ik}=[\boldsymbol{\Phi}]_{ki} =q_iq_k\phi_{ik}\breve{\sigma}_{i}\breve{\sigma}_{k}$, and $\phi_{ik}\in(-1, 1)$, for $i, k=1, 2, 3$ and $i\neq k$. Then, the vector
\begin{align}
\boldsymbol{\Lambda}_\mathbf{q} \frac{\displaystyle\int_{\mathbb{R}_+^3}\mathbf{u}\exp\left\{-\mathbf{u}^\Trans\boldsymbol{\Phi}\mathbf{u}\right\}{\rm d}\mathbf{u}}{\displaystyle\int_{\mathbb{R}_+^3}\exp\left\{-\mathbf{u}^\Trans\boldsymbol{\Phi}\mathbf{u}\right\}{\rm d}\mathbf{u}}
\label{eqn: key_expression_in_lemma2_for_theorem}
\end{align}
is nonlinear in $\mathbf{q}$ if $\boldsymbol{\Phi}$ is a full matrix with all non-zero elements or if exactly one of the following three equalities holds: $\phi_{12}=0$, $\phi_{13}=0$, or $\phi_{23}=0$. 
\end{lemma}

\begin{IEEEproof}
    See Appendix~\ref{app: loss_of_linearity_N3}.
\end{IEEEproof}

\smallskip

\begin{remark}\label{remark: full matrix}
In Lemma~\ref{lemma: OrthantP_3_nonL}, \eqref{eqn: key_expression_in_lemma2_for_theorem} is linear in $\mathbf{q}$ if $\boldsymbol{\Phi}$ is diagonal or if one of the following three conditions holds: $\phi_{12}=\phi_{13}=0$, $\phi_{12}=\phi_{23}=0$, or $\phi_{13}=\phi_{23}=0$. In other words,  to ensure the linearity of \eqref{eqn: key_expression_in_lemma2_for_theorem}  in $\mathbf{q}$,  $\boldsymbol{\Phi}$ must indicate at most pair-wise correlation. 
\end{remark}

\begin{theorem} \label{theo: Theorem_BLMMSE_opt}
Assuming flat Rayleigh fading, spatially and temporally white Gaussian noise, and one-bit ADCs, the BLMMSE channel estimator is MSE-optimal if and only if  each row (resp. each column) of $\mathbf{C}$ in \eqref{eqn: Matrix_C} contains at most two non-zero elements.
\end{theorem}

\begin{IEEEproof}
    See Appendix~\ref{App: proof_Theorem_BLMMSE_opt}. The proof hinges on Lemma~\ref{lemma: OrthantP_2_linearity} and Lemma~\ref{lemma: OrthantP_3_nonL} for the forward part and the converse, respectively.
\end{IEEEproof}

\smallskip

According to Theorem~\ref{theo: Theorem_BLMMSE_opt}, for the BLMMSE channel estimator to be MSE-optimal, the covariance-type matrix $\mathbf{C}$ in \eqref{eqn: Matrix_C} can indicate at most pair-wise correlation. One can relate this condition to Remark~\ref{remark: full matrix}. Through the structure of $\mathbf{C}$ in \eqref{eqn: Matrix_C}, we see that the optimality of the BLMMSE channel estimator is determined not only by the channel correlation, but also by the structure of the pilot matrix. For example, consider the case without channel correlation, i.e., $\boldsymbol{\Sigma}=\mathbf{I}_{N_TN_R}$, for which we have $\boldsymbol{\Omega}_{\rm b}= (\mathbf{S}\mathbf{S}^\Herm+\sigma^2\mathbf{I}_\tau)\otimes\mathbf{I}_{N_R}$. In this case, whether the condition in Theorem~\ref{theo: Theorem_BLMMSE_opt} holds solely depends on the structure of the pilot matrix.

The condition in Theorem~\ref{theo: Theorem_BLMMSE_opt} is rather restrictive, which implies that  $\hat{\mathbf{h}}_{\rm BLM}$ is generally suboptimal in terms of MSE. Nevertheless, in the following we present cases where $\hat{\mathbf{h}}_{\rm MMSE}$ is directly computed to be equivalent to $\hat{\mathbf{h}}_{\rm BLM}$.  All of these cases corroborate Theorem~\ref{theo: Theorem_BLMMSE_opt}.

\subsection{Spatially White Channel and Unitary Pilot Matrix} \label{sec_results: Uncorr_CH_Ortho_Pilot_mat}

Here we assume $\boldsymbol{\Sigma}=\mathbf{I}_{N_TN_R}$ and  a unitary pilot matrix with $\tau=N_T$. Thus,  $\mathbf{S}\mathbf{S}^\Herm=\eta\mathbf{I}_{\tau}$ (with $\eta > 0$), $\mathbf{D}_{\mathsf{I}} = \mathbf{0}$,   and 
\begin{align}
\mathbf{D}_{\mathsf{R}}=\frac{1}{\eta+\sigma^2}\mathbf{I}_{\tau N_R} =\boldsymbol{\Lambda}_{\mathsf{R}}\mathbf{D}_{\mathsf{R}}\boldsymbol{\Lambda}_{\mathsf{R}} = \boldsymbol{\Lambda}_{\mathsf{I}}\mathbf{D}_{\mathsf{R}}\boldsymbol{\Lambda}_{\mathsf{I}}=\boldsymbol{\Omega}_{\rm b}^{-1}.
\end{align}
Clearly, $\mathbf{C}=\frac{1}{\eta+\sigma^2}\mathbf{I}_{2\tau N_R} $, which satisfies the condition in Theorem~\ref{theo: Theorem_BLMMSE_opt}. Since $\mathbf{D}_{\mathsf{I}} = \mathbf{0}$, we can use the expression for $\hat{\mathbf{h}}_{\rm MMSE}$ in \eqref{eqn: special_h_MMSE_D_I=0} (see Appendix~\ref{App: special_case_h_MMSE}). Based on \eqref{eqn: orthantP_diag_mat}, \eqref{eqn: special_h_MMSE_D_I=0}, and 
\eqref{eqn: full_BLMMSE_formula}, we obtain
\begin{align}
\hat{\mathbf{h}}_{\rm MMSE}= \frac{(\mathbf{S}^\Herm\otimes\mathbf{I}_{N_R})\mathbf{r}}{\sqrt{\pi(\eta+\sigma^2)}}=\hat{\mathbf{h}}_{\rm BLM}. \label{eqn: spatially_white_unitary_h_MMSE}
\end{align}

\begin{remark}
Based on \eqref{eqn: spatially_white_unitary_h_MMSE}, for a spatially white Rayleigh fading channel and a unitary pilot matrix, the low-complexity BLMMSE channel estimator is MSE-optimal and thus its use is justified for low spatial correlation. A similar conclusion was reached in~\cite{Utschick_etal_2023}, but only for the spatially white SIMO case with a single pilot symbol.  \label{remark: justification of BLMMSE}
\end{remark}

Next, we generalize the above result and showcase the interplay between the choice of the pilot matrix and the channel correlation.  

\subsection{Transmit-Only Channel Correlation}\label{sec_results:Corr_TX_EVD}

Let $\tau=N_T$. Assume that $\mathbf{H}=\mathbf{H}_{\rm w}\boldsymbol{\Sigma}_{\rm TX}^{\Trans/2}$, where the elements of $\mathbf{H}_{\rm w}$ are i.i.d. $\mathcal{CN}(0, 1)$ random variables. Then, $\mathbf{h}=\VEC(\mathbf{H})\sim\mathcal{CN}(\mathbf{0}, \boldsymbol{\Sigma})$, with $\boldsymbol{\Sigma}=\boldsymbol{\Sigma}_{\rm TX}\otimes\mathbf{I}_{N_R}$. Here, $\boldsymbol{\Omega}_{\rm b}$ in \eqref{eqn: corr_mat_of_b} can be expressed as
\begin{align}
    \boldsymbol{\Omega}_{\rm b}=(\mathbf{S}\boldsymbol{\Sigma}_{\rm TX}\mathbf{S}^\Herm)\otimes\mathbf{I}_{N_R}+\sigma^2\mathbf{I}_{N_TN_R}, \label{eqn: Tx_corr_motivation}
\end{align}
which can be made diagonal by choosing $\mathbf{S}$ according to $\boldsymbol{\Sigma}_{\rm TX}$. To this end, denote the eigenvalue decomposition of $\boldsymbol{\Sigma}_{\rm TX}$ as 
\begin{align}
\boldsymbol{\Sigma}_{\rm TX}=\mathbf{U}\boldsymbol{\Xi}\mathbf{U}^\Herm, \quad \boldsymbol{\Xi}=\Diag{}\big(\left[\xi_i\right]_{i=1}^{N_T}\big),
\label{eqn: Sigma_TX_corr_EVD}
\end{align}
with $\xi_i>0$, $\forall i$.  Now, choosing 
$\mathbf{S}=\sqrt{\eta}\mathbf{U}^\Herm$ (with $\eta>0$) yields
\begin{align}
\boldsymbol{\Omega}_{\rm b} =(\eta\boldsymbol{\Xi} +\sigma^2\mathbf{I}_{N_T})\otimes\mathbf{I}_{N_R}, & \quad
\mathbf{D}_{\mathsf{I}}=\mathbf{0},
\\\mathbf{D}_{\mathsf{R}} =(\eta\boldsymbol{\Xi} +\sigma^2\mathbf{I}_{N_T})^{-1}\otimes\mathbf{I}_{N_R}, & \quad
\mathbf{C}=\BDiag(\mathbf{D}_{\mathsf{R}}, \mathbf{D}_{\mathsf{R}}).
\end{align}
Here $\mathbf{C}$ satisfies the condition in Theorem~\ref{theo: Theorem_BLMMSE_opt}. Since $\mathbf{D}_{\mathsf{I}} = \mathbf{0}$,  \eqref{eqn: special_h_MMSE_D_I=0} holds. From \eqref{eqn: orthantP_diag_mat}, \eqref{eqn: special_h_MMSE_D_I=0}, and
\eqref{eqn: full_BLMMSE_formula}, we obtain
\begin{align}
&\hat{\mathbf{h}}_{\rm MMSE}   = \hat{\mathbf{h}}_{\rm BLM} \nonumber
\\& =\left(\left(\mathbf{U}\;\Diag{}\left(\left[\frac{\xi_i\sqrt{\eta}}{\sqrt{\eta\xi_i+\sigma^2}} \right]_{i=1}^{N_T}\right)\right)\otimes\mathbf{I}_{N_R}\right)\frac{\mathbf{r}}{\sqrt{\pi}}.\label{eqn: TX_corr_equivalence}
\end{align}

Clearly, \eqref{eqn: spatially_white_unitary_h_MMSE} is a special case of \eqref{eqn: TX_corr_equivalence} with $\boldsymbol{\Sigma}_{\rm TX}=\mathbf{I}_{N_T}$. In addition, if $\boldsymbol{\Sigma}_{\rm TX}$ is a circulant matrix~\cite{R_M_Gray}, then a Fourier matrix can be used as the pilot matrix.

From the above, we can see the joint effect of transmit channel correlation and pilot design on the structure of the MMSE channel estimate. Note that, for a general channel covariance matrix  $\boldsymbol{\Sigma}$, the pilot design cannot guarantee that $\hat{\mathbf{h}}_{\rm MMSE}$ is linear.

\begin{remark}\label{remark: TX_corr} 
Assume a standardized $\boldsymbol{\Sigma}_{\rm TX}$ (cf. \eqref{eqn: standardized_cov}) with  $\mathrm{tr}(\boldsymbol{\Sigma}_{\rm TX})=\mathrm{tr}(\boldsymbol{\Xi})=N_T$. Using \eqref{eqn: MSE_per_antenna_general}, the per-antenna MSE corresponding to both estimators in \eqref{eqn: TX_corr_equivalence}  is given by
\begin{align}
       \textnormal{MSE}^{\rm BLM} & =1-\frac{2}{\pi N_T} \sum_{i=1}^{N_T}\frac{\eta\xi_i^2}{\eta\xi_i+\sigma^2}. \label{eqn: MSE_PA_TX_CORR}
\end{align}
Clearly, we have
\begin{align}
\lim_{\sigma^2\rightarrow 0} \textnormal{MSE}^{\rm BLM} &=  1-\frac{2}{\pi},
      \end{align}
which is also valid for \eqref{eqn: spatially_white_unitary_h_MMSE}~\cite{Y_Li_et_al_BLMMSE}. 
\end{remark}

\subsection{Spatially White Channel and $\tau=2$ Real-Valued Pilot Symbols} \label{sec_results:Uncorr_CH_2real_pilot_vec}

Let $\boldsymbol{\Sigma}=\mathbf{I}_{N_T N_R}$ and $\tau=2$, and assume $\mathbf{S}^\Trans = [\mathbf{s}_1 \ \mathbf{s}_2]$ with $\mathbf{s}_1, \mathbf{s}_2 \in \mathbb{R}^{N_T}$. Clearly,  we have $\boldsymbol{\Omega}_{\rm b} = \left(\mathbf{S}\mathbf{S}^\Trans+\sigma^2\mathbf{I}_2\right)\otimes\mathbf{I}_{N_R}$ and, since $\mathbf{D}_{\mathsf{I}} = \mathbf{0}$, $\mathbf{C}=\BDiag(\boldsymbol{\Lambda}_{\mathsf{R}}\mathbf{D}_{\mathsf{R}}\boldsymbol{\Lambda}_{\mathsf{R}}, \boldsymbol{\Lambda}_{\mathsf{I}}\mathbf{D}_{\mathsf{R}}\boldsymbol{\Lambda}_{\mathsf{I}})$. Define
\begin{align}
\boldsymbol{\Delta}_{\rm R} & =\Diag{\big(\left[r_{\mathsf{R},i}r_{\mathsf{R},N_R+i}\right]_{i=1}^{N_R}\big)}, \\
\boldsymbol{\Delta}_{\rm I} & =\Diag{\big(\left[r_{\mathsf{I},i}r_{\mathsf{I},N_R+i}\right]_{i=1}^{N_R}\big)}, 
\end{align}
 and 
 \begin{align}\mathbf{s}_{i, {\rm new}}&=\frac{\mathbf{s}_i}{\sqrt{\|\mathbf{s}_i\|^2+\sigma^2}}, \quad
 \textrm{for}~i=1, 2, \label{eqn: s1_s2_tau_2}
\\\mathbf{S}_{\rm w1}^\Trans &=\left[\mathbf{s}_{1, {\rm new}} \ \mathbf{s}_{2, {\rm new}}\right], \quad \gamma_{s}=\mathbf{s}_{1, {\rm new}}^\Trans\mathbf{s}_{2, {\rm new}}.
\end{align} 
It can be readily shown that
\begin{align}
\boldsymbol{\Lambda}_{\mathsf{R}}\mathbf{D}_{\mathsf{R}}\boldsymbol{\Lambda}_{\mathsf{R}}&=\frac{\begin{bmatrix}
(\|\mathbf{s}_2\|^2+\sigma^2)\mathbf{I}_{N_R} &-\mathbf{s}_1^\Trans\mathbf{s}_2\boldsymbol{\Delta}_{\rm R} \\-\mathbf{s}_1^\Trans\mathbf{s}_2\boldsymbol{\Delta}_{\rm R} & (\|\mathbf{s}_1\|^2+\sigma^2)\mathbf{I}_{N_R}\end{bmatrix}}{\left|\mathbf{S}\mathbf{S}^\Trans+\sigma^2\mathbf{I}_2\right|}, \label{eqn: C_added_real_part}
\end{align}
and  $\boldsymbol{\Lambda}_{\mathsf{I}}\mathbf{D}_{\mathsf{R}}\boldsymbol{\Lambda}_{\mathsf{I}}$ is obtained by replacing $\boldsymbol{\Delta}_{\rm R}$ with $\boldsymbol{\Delta}_{\rm I}$ in \eqref{eqn: C_added_real_part}. Therefore, the structure of $\mathbf{C}$ clearly satisfies the condition in Theorem~\ref{theo: Theorem_BLMMSE_opt}. 

Since $\mathbf{D}_{\mathsf{I}} = \mathbf{0}$, we can use  \eqref{eqn: special_h_MMSE_D_I=0}. It can be shown that, for $k=1, \ldots, N_R$, we have
\begin{align}
\frac{\mathcal{P}\big(\big((\boldsymbol{\Lambda}_{\mathsf{R}}\mathbf{D}_{\mathsf{R}}\boldsymbol{\Lambda}_{\mathsf{R}})_{-k}\big)^{-1}\big)}
{\mathcal{P}(\boldsymbol{\Lambda}_{\mathsf{R}}\boldsymbol{\Omega}_{\rm b}\boldsymbol{\Lambda}_{\mathsf{R}})}
& =\frac{\mathcal{P}\big(\big((\boldsymbol{\Lambda}_{\mathsf{R}}\mathbf{D}_{\mathsf{R}}\boldsymbol{\Lambda}_{\mathsf{R}})_{-(k+N_R)}\big)^{-1}\big)}
{\mathcal{P}(\boldsymbol{\Lambda}_{\mathsf{R}}\boldsymbol{\Omega}_{\rm b}\boldsymbol{\Lambda}_{\mathsf{R}})} \nonumber
\\&=\frac{2\big(1-\frac{r_{\mathsf{R},k}r_{\mathsf{R},N_R+k}\arcsin{\gamma_s}}{\pi/2}\big)}{1-\big(\frac{\arcsin{\gamma_s}}{\pi/2}\big)^2}. \label{eqn: added_equivalence_intermediate}
   \end{align}
Inserting \eqref{eqn: added_equivalence_intermediate} into \eqref{eqn: special_h_MMSE_D_I=0}, after some algebra we obtain
\begin{align}
&\hat{\mathbf{h}}_{\rm MMSE}=\hat{\mathbf{h}}_{\rm BLM}\nonumber
\\
&=\left(\left(\mathbf{S}_{\rm w1}^\Trans
\begin{bmatrix}
1 &\frac{\arcsin\left(\gamma_{s}\right)}{\pi/2} \\
\frac{\arcsin\left(\gamma_{s}\right)}{\pi/2} &1
\end{bmatrix}^{-1}\right)\otimes \mathbf{I}_{N_R}\right)\frac{\mathbf{r}}{\sqrt{\pi}}.\label{eqn: equivalence_2_real_pilots} 
\end{align}

\subsection{Spatially White Channel and $\tau=2$ Complex-Valued Pilot Symbols with $\pi/2$ Phase Difference} \label{sec_results: Uncorr_CH_half_pi_phase}

Let $\boldsymbol{\Sigma}=\mathbf{I}_{N_TN_R}$ and $\tau=2$, and assume $\mathbf{S}^\Trans=[\mathbf{s}_1 \ \mathbf{s}_2]$ with
\begin{align}
\frac{\mathbf{s}_2}{\|\mathbf{s}_2\|} = j \frac{\mathbf{s}_1}{\|\mathbf{s}_1\|}.
\end{align}
Then, it can be shown that
\begin{align}
\boldsymbol{\Omega}_{\rm b}  &= \begin{bmatrix}
\|\mathbf{s}_1\|^2+\sigma^2 &-j\|\mathbf{s}_1\|\|\mathbf{s}_2\|
\\j\|\mathbf{s}_1\|\|\mathbf{s}_2\| &\|\mathbf{s}_2\|^2+\sigma^2
\end{bmatrix}\otimes\mathbf{I}_{N_R}, \\\mathbf{D}_{\mathsf{I}}  &= \frac{\|\mathbf{s}_1\|\|\mathbf{s}_2\|\begin{bmatrix}
\mathbf{0} &\mathbf{I}_{N_R}
\\-\mathbf{I}_{N_R} &\mathbf{0}
\end{bmatrix}}{\sigma^2\big(\sigma^2+\mathrm{tr}(\mathbf{S}\mathbf{S}^\Herm)\big)},
\\\mathbf{D}_{\mathsf{R}}  &= \frac{\mathbf{C}_{{\rm sub, A}} } {\sigma^2\big(\sigma^2+\mathrm{tr}(\mathbf{S}\mathbf{S}^\Herm)\big)}, 
\end{align}
with
\begin{align}
\mathbf{C}_{{\rm sub, A}} & = \BDiag{}\left((\|\mathbf{s}_2\|^2+\sigma^2)\mathbf{I}_{N_R}, \big(\|\mathbf{s}_1\|^2+\sigma^2\big)\mathbf{I}_{N_R}\right).
\end{align}
 Moreover, by defining
\begin{align}
\boldsymbol{\Delta}_{\rm X} & =\Diag{\big(\left[r_{\mathsf{R},i}r_{\mathsf{I},N_R+i}\right]_{i=1}^{N_R}\big)}, \\
\boldsymbol{\Delta}_{\rm Y} & =\Diag{\big(\left[r_{\mathsf{I},i}r_{\mathsf{R},N_R+i}\right]_{i=1}^{N_R}\big)},  \\
\mathbf{C}_{{\rm sub, B}} & =\begin{bmatrix}
\mathbf{0} & -\|\mathbf{s}_1\|\|\mathbf{s}_2\|\boldsymbol{\Delta}_{\rm X}\\\|\mathbf{s}_1\|\|\mathbf{s}_2\|\boldsymbol{\Delta}_{\rm Y}&\mathbf{0} 
\end{bmatrix},
\end{align}
we have
\begin{align}
\mathbf{C}=\frac{1}{\sigma^2\left(\sigma^2+\mathrm{tr}(\mathbf{S}\mathbf{S}^\Herm)\right)}\begin{bmatrix}
\mathbf{C}_{\rm sub, A} & \mathbf{C}_{\rm sub, B} \\
\mathbf{C}_{\rm sub, B}^\Trans
&\mathbf{C}_{\rm sub, A}
\end{bmatrix},
\end{align}
which clearly satisfies the condition in Theorem~\ref{theo: Theorem_BLMMSE_opt}. 
Reuse \eqref{eqn: s1_s2_tau_2} in the new context here and denote 
\begin{align}
\mathbf{S}_{\rm w2}^\Herm = \left[\mathbf{s}_{1, {\rm new}}^* \ \mathbf{s}_{2, {\rm new}}^*\right], \quad
\widetilde{\gamma}_{s}= \|\mathbf{s}_{1, {\rm new}}\|\|\mathbf{s}_{2, {\rm new}}\|.
\end{align}
Using similar techniques as in Section~\ref{sec_results:Uncorr_CH_2real_pilot_vec}, from  \eqref{eqn: reduction_2}--\eqref{eqn: new_g_k} and \eqref{eqn: full_BLMMSE_formula}, we obtain 
\begin{align}
&\hat{\mathbf{h}}_{\rm MMSE}=\hat{\mathbf{h}}_{\rm BLM}
\nonumber
\\&=\left(\left(\mathbf{S}_{\rm w2}^\Herm
\begin{bmatrix}
1 &\frac{-j\arcsin\left(\widetilde{\gamma}_{s}\right)}{\pi/2}
\\\frac{j\arcsin\left(\widetilde{\gamma}_{s}\right)}{\pi/2} &1
\end{bmatrix}^{-1}\right)\otimes \mathbf{I}_{N_R}\right)\frac{\mathbf{r}}{\sqrt{\pi}}.\label{eqn: equivalence_complex_tau2}
\end{align}

\begin{corollary}\label{corollary: QPSK}
With $N_T=1$, $\tau=2$, and arbitrary $N_R$, $\hat{\mathbf{h}}_{\rm BLM}$ and $\hat{\mathbf{h}}_{\rm MMSE}$ are equivalent for spatially white channels if quadrature phase-shift keying symbols are used as pilot symbols. 
\end{corollary}
\begin{IEEEproof}
If the phase difference of the two pilot symbols $s_1$ and $s_2$ is $\pi$, the result in Section~\ref{sec_results:Uncorr_CH_2real_pilot_vec} applies. If the phase difference is $\frac{\pi}{2}$, the result in this section applies. In both cases, the condition in Theorem~\ref{theo: Theorem_BLMMSE_opt} is satisfied.
\end{IEEEproof}

\smallskip

The pilot symbols in Section~\ref{sec_results:Uncorr_CH_2real_pilot_vec} and this section are chosen to ensure the equivalence between $\hat{\mathbf{h}}_{\rm MMSE}$ and $\hat{\mathbf{h}}_{\rm BLM}$ only. Designing the pilot matrix to minimize the MSE is interesting but beyond the scope of this work. We refer to~\cite{Atzeni_2022} for more details on this subject.

\subsection{SIMO with Real-Valued Channel Correlation, $N_R=2$, and $\tau=N_T=1$} \label{sec_results: Fully_Corr_N2}

The nonlinearity of the MMSE channel estimate for noiseless correlated SIMO channels was considered in~\cite{Utschick_etal_2023}. Here, we focus on a general noisy correlated SIMO channel with a single complex-valued pilot symbol (i.e., $\tau=N_T=1$). In this setting, the system model in \eqref{sm: vec_SM} becomes
\begin{align}
\mathbf{y}= \mathbf{h} s +\mathbf{n} \in \mathbb{C}^{N_R}. \label{eqn: SIMO_1}
\end{align}
 Note that, for a SIMO channel, $\boldsymbol{\Sigma}$ represents the correlation among receive antennas. Assume that $\boldsymbol{\Sigma}$ is real-valued, e.g., as in the exponential channel model~\cite{Valentine_Aalo_exp_model,Chiani_exp_model}. Then,
\begin{align}
\mathbf{D}_{\mathsf{R}}^{-1} =\boldsymbol{\Omega}_{\rm b}=|s|^2\boldsymbol{\Sigma}+\sigma^2\mathbf{I}_{N_R}, \quad \mathbf{D}_{\mathsf{I}}=\mathbf{0}. \label{eqn: SIMO_2}
\end{align}
 Hence, \eqref{eqn: special_h_MMSE_D_I=0} in Appendix~\ref{App: special_case_h_MMSE} holds. For the case with $N_R=2$, since $\mathbf{C}=\BDiag(\boldsymbol{\Lambda}_{\mathsf{R}}\mathbf{D}_{\mathsf{R}}\boldsymbol{\Lambda}_{\mathsf{R}}, \boldsymbol{\Lambda}_{\mathsf{I}}\mathbf{D}_{\mathsf{R}}\boldsymbol{\Lambda}_{\mathsf{I}})$, the condition in Theorem~\ref{theo: Theorem_BLMMSE_opt} holds trivially. Assume a standardized $\boldsymbol{\Sigma}$, with channel correlation coefficient given by  $\rho_{12}$, and define $\beta_{12}= \frac{\rho_{12}|s|^2}{|s|^2+\sigma^2}$. Based on \eqref{eqn: OrthantP_N2},  \eqref{eqn: special_h_MMSE_D_I=0} with $\mathbf{A}=s\mathbf{I}_2$, and \eqref{eqn: full_BLMMSE_formula},  we obtain
\begin{align}
\hat{\mathbf{h}}_{\rm MMSE} = \frac{s^*\boldsymbol{\Sigma}\begin{bmatrix}
1 & \frac{\arcsin(\beta_{12})}{\pi/2} \\
\frac{\arcsin(\beta_{12})}{\pi/2} & 1
\end{bmatrix}^{-1}\mathbf{r}}{\sqrt{\pi(|s|^2+\sigma^2)}}
=\hat{\mathbf{h}}_{\rm BLM}. \label{eqn: SIMO_N_eq_2}
\end{align}

\section{Cases with Nonlinear MMSE Channel Estimators}\label{sec: non-L results}

We now consider two cases where the  condition in Theorem~\ref{theo: Theorem_BLMMSE_opt} does not hold and, thus, $\hat{\mathbf{h}}_{\rm MMSE}$ is nonlinear in $\mathbf{r}$. For general MIMO systems with arbitrary channel correlation and pilot symbols, one can always resort to our explicit expressions in \eqref{eqn: reduction_2}--\eqref{eqn: new_g_k} and use existing methods discussed in Section~\ref{Sec: OrthantP_basics} (e.g., the algorithm for Monte Carlo simulations in~\cite{Genz_algorithm}) to compute the required orthant probabilities for one covariance matrix of dimension $(2\tau N_R)\times (2\tau N_R)$ and $2\tau N_R$ covariance matrices of dimension $(2\tau N_R-1)\times (2\tau N_R-1)$. Clearly, when either $\tau$ or $N_R$ is large, Monte Carlo methods become computationally cumbersome and cannot guarantee sufficient accuracy. Nevertheless, using our insights above (e.g., Remark~\ref{remark: on pairwise_quantized_op_in_C}) and existing results on the Gaussian orthant probability, computationally efficient analytical expressions for $\hat{\mathbf{h}}_{\rm MMSE}$ can be derived for various SIMO channels  under specific assumptions on the channel correlation or pilot symbols. 

\subsection{SIMO Channel with Real-Valued Channel Correlation,  $N_R\geq 3$, and $\tau=N_T=1$} \label{sec: SIMO_tau_1}

Consider the same  SIMO system described by \eqref{eqn: SIMO_1} in Section~\ref{sec_results: Fully_Corr_N2} with $\tau=N_T=1$ and  $N_R\geq 3$. In this case,  \eqref{eqn: SIMO_2} remains applicable. Without loss of generality, assume a standardized $\boldsymbol{\Sigma}$ with channel correlation coefficients  denoted by $\rho_{ik}$, for $i, k = 1, \ldots, N_R$ and $i\neq k$, unless otherwise stated.

\begin{theorem}\label{theo: SIMO_N3_nonL}
Consider the one-bit quantized, spatially correlated Rayleigh fading SIMO system given by \eqref{eqn: SIMO_1} with $\tau=N_T=1$ and $N_R=3$. 
Then,
\begin{align}
\label{eqn: h_mmse_N3}
\hat{\mathbf{h}}_{\rm MMSE}= \frac{s^*\boldsymbol{\Sigma}}{2\sqrt{\pi \left(|s|^2+\sigma^2\right)}} \left(\frac{\breve{\mathbf{v}}_\mathsf{R}}{\breve{P}_\mathsf{R}}+\frac{j\breve{\mathbf{v}}_{\mathsf{I}}}{\breve{P}_{\mathsf{I}}}\right),
\end{align}
with 
\begin{align}
\beta_{ik} & = \frac{\rho_{ik}|s|^2}{|s|^2+\sigma^2} \quad (i, k=1, 2, 3, \; i\neq k), \label{eqn: tilde_beta_ij} \\
\breve{\mathbf{v}}_\mathsf{R} & = \begin{bmatrix}
\frac{r_{\mathsf{R},1}}{4}+\frac{r_{\mathsf{R},1}r_{\mathsf{R},2}r_{\mathsf{R},3}}{2\pi}\arcsin\left(\frac{\beta_{23}-\beta_{12}\beta_{13}}{\sqrt{1-\beta_{12}^2}\sqrt{1-\beta_{13}^2}}\right)
\\ \frac{r_{\mathsf{R},2}}{4}+\frac{r_{\mathsf{R},1}r_{\mathsf{R},2}r_{\mathsf{R},3}}{2\pi}\arcsin\left(\frac{\beta_{13}-\beta_{12}\beta_{23}}{\sqrt{1-\beta_{12}^2}\sqrt{1-\beta_{23}^2}}\right)\\ \frac{r_{\mathsf{R},3}}{4}+\frac{r_{\mathsf{R},1}r_{\mathsf{R},2}r_{\mathsf{R},3}}{2\pi}\arcsin\left(\frac{\beta_{12}-\beta_{13}\beta_{23}}{\sqrt{1-\beta_{13}^2}\sqrt{1-\beta_{23}^2}}\right)
\end{bmatrix}, \label{eqn: mean_vec_h_MMSE_SIMO_N3}
\\\breve{P}_\mathsf{R} & = \frac{1}{8}+\frac{\sum_{i=1}^2\sum_{k=i+1}^3r_{\mathsf{R},i}r_{\mathsf{R},k}\arcsin( \beta_{ik})}{4\pi}.
\label{eqn: ortP_R_h_MMSE_SIMO_N3}
\end{align}
Note that $\breve{\mathbf{v}}_\mathsf{I}$ (resp. $\breve{P}_\mathsf{I}$) in \eqref{eqn: h_mmse_N3} is the same as $\breve{\mathbf{v}}_\mathsf{R}$ (resp. $\breve{P}_\mathsf{R}$) except that $r_{\mathsf{R}, i}$ in \eqref{eqn: mean_vec_h_MMSE_SIMO_N3} (resp. \eqref{eqn: ortP_R_h_MMSE_SIMO_N3}) is replaced by $r_{\mathsf{I}, i}$, for $i=1, 2, 3$.
\end{theorem}

\begin{IEEEproof}
The proof starts from \eqref{eqn: special_h_MMSE_D_I=0} with $\mathbf{A}=s\mathbf{I}_3$ and can be obtained from \eqref{eqn: OrthantP_identity}--\eqref{eqn: OrthantP_scaling}, \eqref{eqn: OrthantP_N2}, and \eqref{eqn: OrthantP_N3}. The details are omitted due to the space limitations.
\end{IEEEproof}

\smallskip

On the other hand, under the same assumptions for Theorem~\ref{theo: SIMO_N3_nonL}, the BLMMSE channel estimate is given by 
\begin{align}
\hat{\mathbf{h}}_{\rm BLM}= \frac{s^*\boldsymbol{\Sigma}\begin{bmatrix}
1 & \frac{\arcsin(\beta_{12})}{\pi/2}&\frac{\arcsin(\beta_{13})}{\pi/2} \\
\frac{\arcsin(\beta_{12})}{\pi/2}&1 & \frac{\arcsin(\beta_{23})}{\pi/2}\\
\frac{\arcsin(\beta_{13})}{\pi/2}& \frac{\arcsin(\beta_{23})}{\pi/2}&1
\end{bmatrix}^{-1}\mathbf{r}}{\sqrt{\pi \left(|s|^2+\sigma^2\right)}}.\label{eqn: h_blmmse_N3}
\end{align}
In general, \eqref{eqn: h_blmmse_N3} is not equivalent to \eqref{eqn: h_mmse_N3} unless  $\boldsymbol{\Sigma}$ is diagonal or if one of the following three conditions holds: $\rho_{23}=\rho_{13}=0$, $\rho_{13}=\rho_{12}=0$, or $\rho_{12}=\rho_{23}=0$. This is consistent with Theorem~\ref{theo: Theorem_BLMMSE_opt}.

\begin{remark}
    Theorem~\ref{theo: SIMO_N3_nonL} is significant in that,  for one-bit quantized multi-antenna systems, it provides the simplest   closed-form expression for $\hat{\mathbf{h}}_{\rm MMSE}$ that demonstrates a nonlinear dependence on $\mathbf{r}$. 
\end{remark}

When $N_R=4$, for the same SIMO system given by \eqref{eqn: SIMO_1} with a real-valued channel covariance matrix, based on \eqref{eqn: OrthantP_N3}, \eqref{eqn: special_h_MMSE_D_I=0} with $\mathbf{A}=s\mathbf{I}_4$, and \eqref{eqn: OrthantP4} in Appendix~\ref{App: OrthantP_basics_P4}, we can obtain a semi-closed-form expression for $\hat{\mathbf{h}}_{\rm MMSE}$ with only the one-dimensional finite integration in \eqref{eqn: OrthantP4}.

Next, we consider the SIMO system in \eqref{eqn: SIMO_1} with arbitrary $N_R$ but with equal positive channel correlation coefficient for all pairs of receive antennas. Let $Q(\cdot)$ denote the Gaussian $Q$-function, i.e., the right tail distribution function of the standard normal distribution $\mathcal{N}(0, 1)$.

\begin{theorem}\label{theo: SIMO_general_N_special_corr}
Consider the one-bit quantized, spatially correlated Rayleigh fading SIMO system given by \eqref{eqn: SIMO_1} with $\tau=N_T=1$. Let the channel correlation coefficients in the standardized $\boldsymbol{\Sigma}$ be given by $\rho_{ik}=\rho$ (with $0\leq \rho \leq 1$), for $i, k=1, \ldots, N_R$ and $i\neq k$. Then,
\begin{align}
\hat{\mathbf{h}}_{\rm MMSE}= \frac{s^*\boldsymbol{\Sigma}}{2\sqrt{\pi\left(|s|^2+\sigma^2\right)}}\left(\frac{\boldsymbol{\Lambda}_{\mathsf{R}}\mathbf{v}_\mathsf{R}}
{P_\mathsf{R}} + \frac{j\boldsymbol{\Lambda}_{\mathsf{I}}\mathbf{v}_\mathsf{I}}
{P_\mathsf{I}}\right), \label{eqn: MMSE_SIMO_same_rho}
\end{align}
with
\begin{align}
P_\mathsf{R} = \displaystyle\int_{-\infty}^\infty \prod_{k=1}^{N_R} Q\left(\frac{r_{\mathsf{R}, k}\sqrt{\rho}|s|t}{\sqrt{|s|^2(1-\rho) +\sigma^2}} \right) \frac{\exp\big\{-\frac{t^2}{2}\big\}}{\sqrt{2\pi}}{\rm d}t. \label{eqn: MMSE_SIMO_same_rho_P_denom}
\end{align}
Note that $P_\mathsf{I}$ in \eqref{eqn: MMSE_SIMO_same_rho} is the same as $P_\mathsf{R}$ except that $r_{\mathsf{R}, k}$ in \eqref{eqn: MMSE_SIMO_same_rho_P_denom} is replaced by $r_{\mathsf{I}, k}$, for $k=1, \ldots, N_R$. The $k$-th element of $\mathbf{v}_\mathsf{R}$ in \eqref{eqn: MMSE_SIMO_same_rho} is given by 
\begin{align}
v_{\mathsf{R}, k} = \displaystyle \int_{-\infty}^\infty \prod_{\substack{i=1\\i\neq k}}^{N_R} Q\left(\frac{r_{\mathsf{R}, i}\sqrt{\rho}|s|t}{\sqrt{|s|^2 +\sigma^2}} \right) \frac{ \exp\big\{-\frac{t^2}{2}\big\}}{\sqrt{2\pi}}{\rm d}t, \label{eqn: MMSE_SIMO_same_rho_P_vec_num}
\end{align}
for $k=1, \ldots, N_R$. Note that the $k$-th element of $\mathbf{v}_\mathsf{I}$, i.e., $v_{\mathsf{I}, k}$, is the same as $v_{\mathsf{R}, k}$, except that $r_{\mathsf{R}, i}$ is replaced by $r_{\mathsf{I}, i}$ in \eqref{eqn: MMSE_SIMO_same_rho_P_vec_num},  for $i=1, \ldots, N_R$ and $i\neq k$.
\end{theorem}

\begin{IEEEproof}
See Appendix~\ref{app: SIMO_MMSE_N_general_Theorem_proof}. The proof utilizes the key observation made in Remark~\ref{remark: on pairwise_quantized_op_in_C} that each off-diagonal element of $\mathbf{C}$ in \eqref{eqn: Matrix_C} is affected by exactly two different elements from $\left[\mathbf{r}_{\mathsf{R}}^\Trans \ \mathbf{r}_{\mathsf{I}}^\Trans\right]^\Trans$. Based on this, the proof further invokes the results in~\cite[p. 192]{Tong_MVN_book}.
\end{IEEEproof}

\smallskip

\begin{remark}
Both \eqref{eqn: MMSE_SIMO_same_rho_P_denom} and \eqref{eqn: MMSE_SIMO_same_rho_P_vec_num} require integration over only one variable. Therefore, \eqref{eqn: MMSE_SIMO_same_rho} can be conveniently evaluated using common computing software  even for large values of $N_R$~\cite{Tong_MVN_book}. This comment also applies to \eqref{eqn: orthant_P_SISO_W_R} and \eqref{eqn: orthantP_SISO_Wx_k} in Theorem~\ref{theo: theo_rank_one_SISO}.
\end{remark}

\subsection{SISO and Spatially White SIMO Channels with Real-Valued Pilot Symbols, Arbitrary $\tau$, and $N_T=1$}

For SIMO systems with spatially white channels, the MMSE channel estimate derived for one receive antenna holds for all receive antennas. This allows us to  simplify the presentation and focus on SISO channels with multiple real-valued pilot symbols. Denote the real-valued pilot vector $\mathbf{s}$  as
\begin{align}
\mathbf{s}=[s_1 \ldots s_\tau]^\Trans.\label{eqn: multi_pilot_vec_real}
\end{align} 
When $\tau=1$, this scenario becomes a special case of the one described in Section~\ref{sec_results: Uncorr_CH_Ortho_Pilot_mat} with $\tau=N_T=N_R=1$. Furthermore, when $\tau=2$, the setup reduces to that in Section~\ref{sec_results:Uncorr_CH_2real_pilot_vec} with $N_T=N_R=1$. Therefore, the non-trivial new case here corresponds to $\tau>2$, although the subsequent results are applicable to $\tau=1, 2$.

When $N_R=N_T=1$, the system model \eqref{sm: vec_SM} reduces to \begin{align}
    \mathbf{y}= h\mathbf{s}+\mathbf{n},
\label{eqn: SISO_sys_model}
\end{align} with $h\sim\mathcal{CN}(0, 1)$ being the scalar channel. Correspondingly, the MMSE (resp. BLMMSE) channel estimate is denoted by $\hat{h}_{\rm MMSE}$ (resp. $\hat{h}_{\rm BLM}$). Furthermore, we have
\begin{align}
\mathbf{D}_{\mathsf{R}}^{-1}=\boldsymbol{\Omega}_{\rm b}=\mathbf{s}\mathbf{s}^\Trans+\sigma^2\mathbf{I}_\tau, \quad \mathbf{D}_{\mathsf{I}}=\mathbf{0}. \label{eqn: SISO_DR_DI_rank_one}
\end{align}
Using the Sherman-Morrison formula~\cite[p. 124]{Meyer_matrix}, we have $\mathbf{D}_{\mathsf{R}}=\frac{1}{\sigma^2}\big(\mathbf{I}_{\tau}-\frac{\mathbf{s}\mathbf{s}^\Trans}{\mathbf{s}^\Trans\mathbf{s}+\sigma^2}\big)$. Correspondingly,
\begin{align}
&\mathbf{D}_{\mathsf{R}, -k}  = \frac{1}{\sigma^2}\bigg(\mathbf{I}_{\tau-1}-\frac{\mathbf{s}_{-k} \mathbf{s}_{-k}^\Trans}{\mathbf{s}^\Trans\mathbf{s}+\sigma^2}\bigg),
\\
&(\mathbf{D}_{\mathsf{R}, -k})^{-1}  = \sigma^2\bigg(\mathbf{I}_{\tau-1}+\frac{\mathbf{s}_{-k} \mathbf{s}_{-k}^\Trans}{s_k^2+\sigma^2}\bigg).\label{eqn: SISO_DRK_inverse}
\end{align}

\begin{theorem}
\label{theo: theo_rank_one_SISO}
  Consider the one-bit quantized Rayleigh fading SISO system in \eqref{eqn: SISO_sys_model} with a real-valued pilot vector given by \eqref{eqn: multi_pilot_vec_real}.  
Then,
\begin{align}
\hat{h}_{\rm MMSE}= \frac{1}{2\sqrt{\pi}}\sum_{k=1}^\tau\frac{s_k}{\sqrt{s_k^2+\sigma^2}}\left(\frac{r_{\mathsf{R}, k}\overline{v}_{\mathsf{R}, k}}{\overline{P}_{\mathsf{R}}}+\frac{jr_{\mathsf{I}, k}\overline{v}_{\mathsf{I}, k}}{\overline{P}_{\mathsf{I}}}\right), \label{eqn: h_MMSE_SISO_multipilot}
\end{align}
with 
\begin{align}
\overline{P}_{\mathsf{R}} & = \displaystyle \int_{-\infty}^\infty \prod_{i=1}^\tau Q\left(\frac{r_{\mathsf{R},i}s_i t}{\sigma}\right)\frac{\exp\big\{-\frac{t^2}{2}\big\}}{\sqrt{2\pi}}{\rm d}t, 
\label{eqn: orthant_P_SISO_W_R} \\
\overline{v}_{\mathsf{R}, k} & = \displaystyle \int_{-\infty}^\infty \prod_{\substack{i=1\\i\neq k}}^\tau Q\left(\frac{r_{\mathsf{R},i}s_i t}{\sqrt{s_k^2+\sigma^2}}\right)\frac{\exp\big\{-\frac{t^2}{2}\big\}}{\sqrt{2\pi}}{\rm d}t, \label{eqn: orthantP_SISO_Wx_k}
\end{align}
for $k=1, \ldots, \tau$. Note that, $\overline{P}_{\mathsf{I}}$ is the same as $\overline{P}_{\mathsf{R}}$ in \eqref{eqn: orthant_P_SISO_W_R} except that $r_{\mathsf{R}, i}$ is replaced by $r_{\mathsf{I}, i}$, for $i=1, \ldots, \tau$. Similarly, $\overline{v}_{\mathsf{I}, k}$ is the same as $\overline{v}_{\mathsf{R}, k}$ in \eqref{eqn: orthantP_SISO_Wx_k} except that $r_{\mathsf{R}, i}$ is replaced by $r_{\mathsf{I}, i}$, for $i = 1, \ldots, \tau$ and $i\neq k$.
\end{theorem}

\begin{IEEEproof}
See Appendix~\ref{app: proof_of_theo_SISO_rank_one}. The main idea is similar to that for the proof of Theorem~\ref{theo: SIMO_general_N_special_corr}.
\end{IEEEproof}

\smallskip

\begin{remark}\label{remark: tau=2_SISO} Let $N_T=N_R=1$. 
As expected, \eqref{eqn: h_MMSE_SISO_multipilot} is equivalent to \eqref{eqn: spatially_white_unitary_h_MMSE}  when $\tau=1$ and to \eqref{eqn: equivalence_2_real_pilots} when $\tau=2$. In both cases, $\hat{h}_{\rm MMSE} = \hat{h}_{\rm BLM}$.  Based on \eqref{eqn: MSE_per_antenna_general}, when $\sigma^2\rightarrow 0$, the  corresponding asymptotic MSE  for both estimators is given by $(1-\frac{2}{\pi})$  with $\tau=2$. In fact,  for $N_T=N_R=1$, $(1-\frac{2}{\pi})$ is the asymptotic MSE   when $\sigma^2\rightarrow 0$ for both \eqref{eqn: equivalence_2_real_pilots} and \eqref{eqn: equivalence_complex_tau2}.
\end{remark}

\section{Numerical Results}\label{sec: simulation results}

In this section, we present numerical results to corroborate the above analyses, where the analytical expressions are verified by means of simulation results. For the cases with a single pilot symbol, the signal-to-noise ratio (SNR) is defined as $\textrm{SNR}=|s|^2/\sigma^2$. For the cases with a $\tau$-dimensional pilot vector $\mathbf{s}$, we have $\textrm{SNR}=\|\mathbf{s}\|^2/(\tau\sigma^2)$. When a pilot matrix is used, we have $\textrm{SNR}=\mathrm{tr}(\mathbf{S}\mathbf{S}^\Herm)/(\tau N_T\sigma^2)$.  The theoretical per-antenna MSE for $\hat{\mathbf{h}}_{\rm BLM}$ is computed using \eqref{eqn: MSE_per_antenna_general}. For the simulation results based on the analytical results (i.e., those in Sections~\ref{sec: BLMMSE_opt_results} and~\ref{sec: non-L results}), the per-antenna MSE is obtained by averaging $\frac{1}{N_TN_R}\|\mathbf{h}-\hat{\mathbf{h}}\|^2$ over at least $10^5$ independent channel realizations, where $\hat{\mathbf{h}}$ can be either $\hat{\mathbf{h}}_{\rm BLM}$ or $\hat{\mathbf{h}}_{\rm MMSE}$. For the simulation results based on the algorithm in~\cite{Genz_algorithm}, at least $10^4$ independent channel realizations are used. Without loss of generality, the channel covariance matrices used here are standardized.

\subsection{MISO and MIMO Channels with Unitary Pilot Matrix and No Receive Correlation}

\begin{figure}[t]
\begin{center}
\includegraphics[width=0.99\columnwidth]
{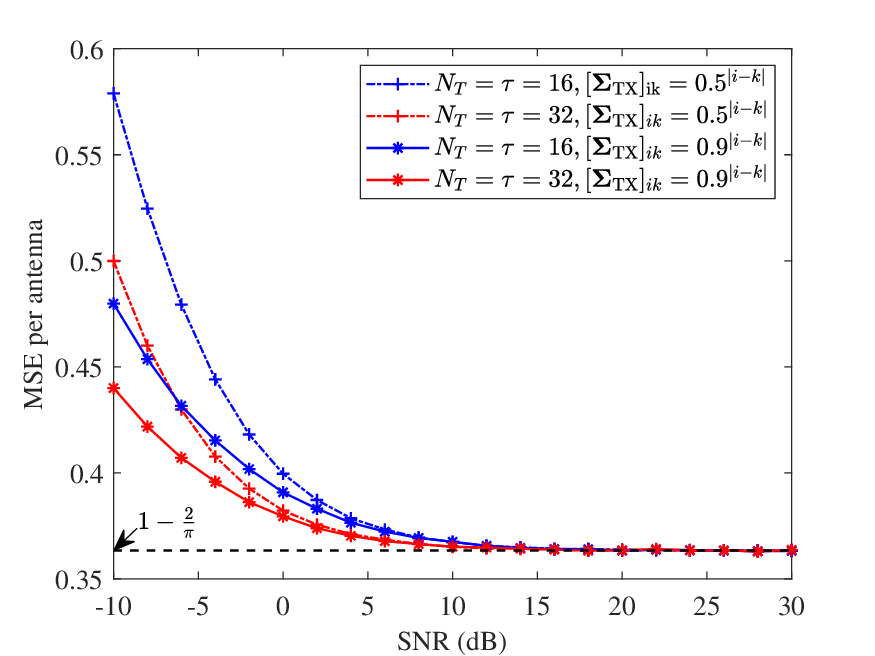}
\end{center}
\caption{MSE comparison for the MISO case with $\tau=N_T \in \{16, 32\}$ using the exponential transmit correlation model and a tailored pilot matrix. In this scenario $\hat{\mathbf{h}}_{\rm MMSE}=\hat{\mathbf{h}}_{\rm BLM}$.}
\label{Fig: MISO_TX_corr}
\end{figure}

 Corresponding to the case studied in Section~\ref{sec_results:Corr_TX_EVD}, here we investigate the impact of transmit antenna correlation and the number of transmit antennas $N_T$.  Given the choice of the pilot matrix for this case, we have $\hat{\mathbf{h}}_{\rm BLM}=\hat{\mathbf{h}}_{\rm MMSE}$. In the absence of channel correlation at the receiver, the MSE is statistically the same for all the receive antennas. Therefore, a MISO setup is considered for simplicity. For the transmit channel correlation in \eqref{eqn: Sigma_TX_corr_EVD}, we use the exponential correlation model~\cite{Chiani_exp_model, Valentine_Aalo_exp_model} and let $[\boldsymbol{\Sigma}_{\rm TX}]_{ik}=0.5^{|i-k|}$ or $[\boldsymbol{\Sigma}_{\rm TX}]_{ik}=0.9^{|i-k|}$, for $i, k=1, \ldots, N_T$. The results are shown in Fig.~\ref{Fig: MISO_TX_corr}. Clearly, increasing both $N_T$ and the amount of transmit correlation improves the MSE performance at low SNR. At high SNR, the per-antenna MSE converges to $(1-\frac{2}{\pi})$, confirming the claim in Remark~\ref{remark: TX_corr}.  Similar trends are observed for other channel correlation models, e.g., those in~\cite{Abdi_2002, Jakes_book}, as reported in~\cite{Ding_et_al_SPAWC2024}.

\subsection{SIMO Channel with $\tau=1$ and $N_R=3,4$} 

\begin{figure}[t]
\begin{center}
\includegraphics[width=0.99\columnwidth]
{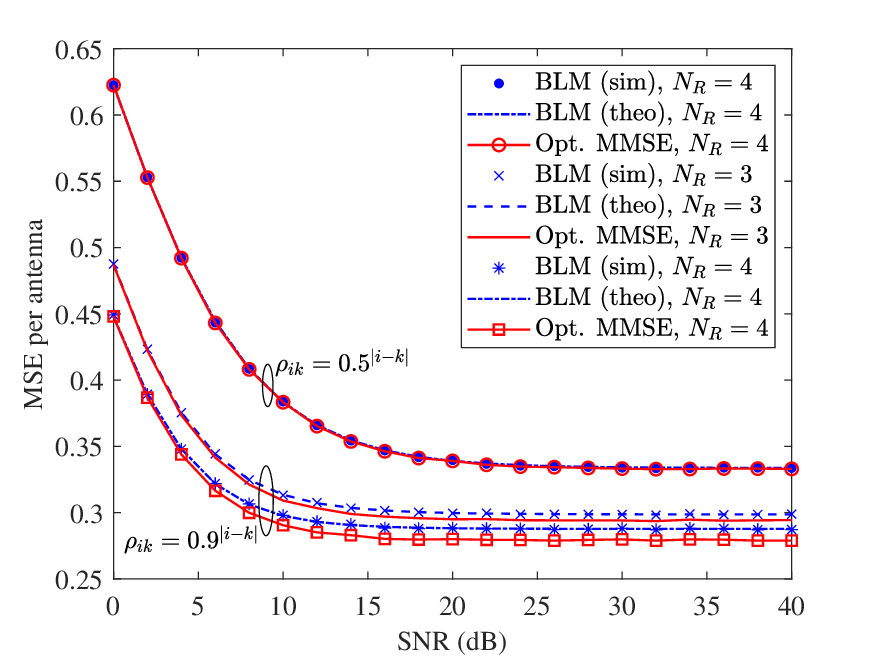}
\end{center}
\caption{MSE comparison for the SIMO case with $N_R \in \{3, 4\}$  using the exponential channel correlation model.}
\label{Fig: SIMO_exp_N34_rho59}
\end{figure}

Here, we assume $N_R=3$ and $N_R=4$ corresponding to the discussion in Section~\ref{sec: SIMO_tau_1}. Fig.~\ref{Fig: SIMO_exp_N34_rho59} shows the results using the exponential channel correlation model~\cite{Valentine_Aalo_exp_model, Chiani_exp_model} with $\rho_{ik}=0.5^{|i-k|}$ or $\rho_{ik}=0.9^{|i-k|}$, for $i, k=1, \ldots, N_R$. We observe that the difference in the MSE performance between $\hat{\mathbf{h}}_{\rm BLM}$ and $\hat{\mathbf{h}}_{\rm MMSE}$ is more pronounced at moderate-to-high SNR when both the channel correlation coefficient and the value of $N_R$ grow large. When the channel correlation is low, the MSE performance of the two estimators is similar as expected according to Theorem~\ref{theo: Theorem_BLMMSE_opt} (see Remark~\ref{remark: justification of BLMMSE}).

\subsection{SIMO Channel with $\tau=1$ and $\rho_{i,k}=\rho$}

\begin{figure}[t]
\begin{center}
\includegraphics[width=0.99\columnwidth]
{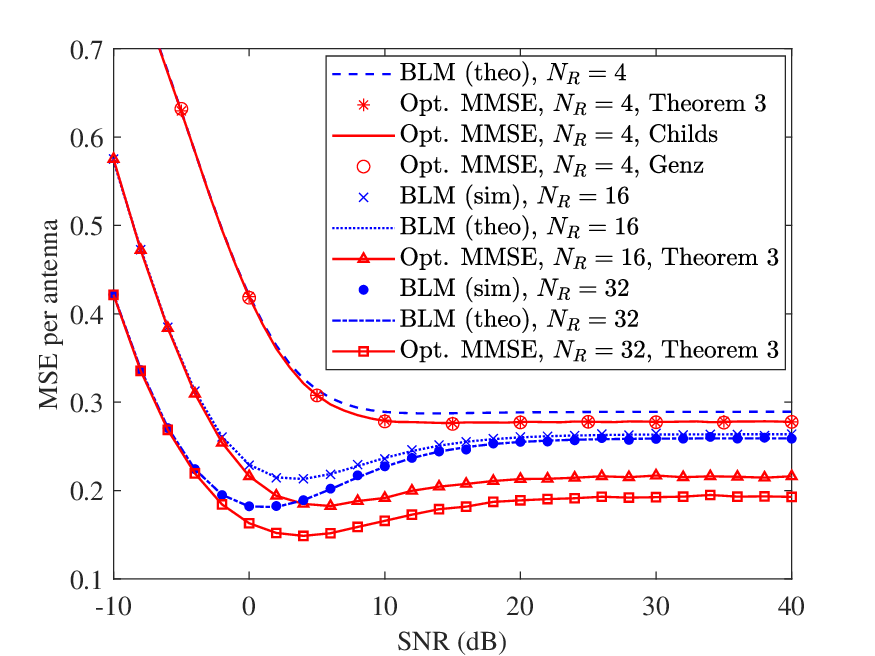}
\end{center}
\caption{MSE comparison for the SIMO case with $N_R \in \{ 4, 16, 32 \}$ and the same channel correlation coefficient $\rho=0.9$ between all antennas.}
\label{Fig: same_cross_coeff_SIMO_N16_32}
\end{figure}

\begin{figure}[t]
\begin{center}
\includegraphics[width=0.99\columnwidth]
{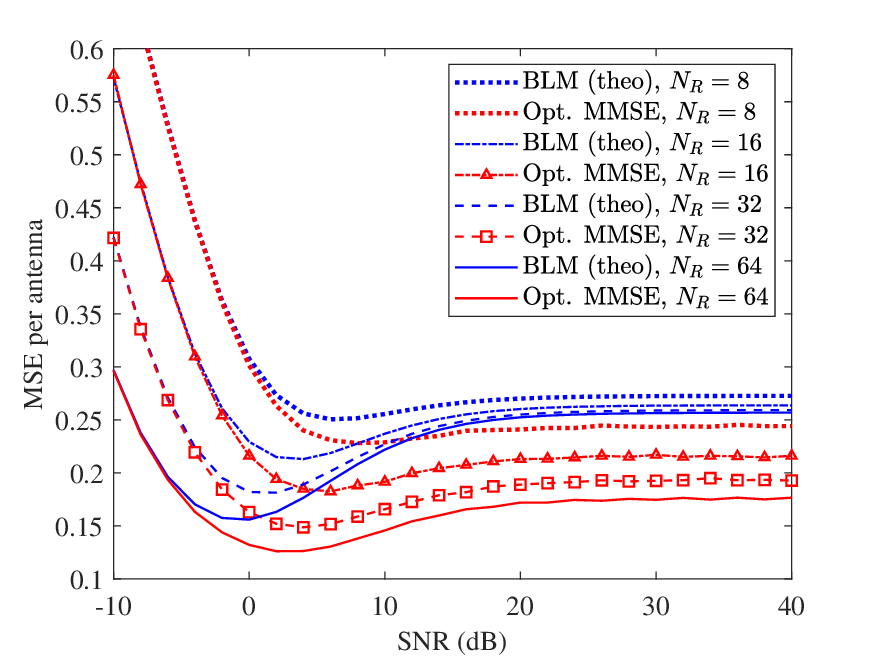}
\end{center}
\caption{MSE comparison for the SIMO case with $N_R \in \{ 8, 16, 32, 64\}$ and the same channel correlation coefficient $\rho=0.9$ between all antennas.}
\label{Fig: same_cross_coeff_SIMO_more_N}
\end{figure}

Using Theorem~\ref{theo: SIMO_general_N_special_corr}, we provide numerical results in Fig.~\ref{Fig: same_cross_coeff_SIMO_N16_32} for $N_R \in \{ 4, 16, 32 \}$ and in Fig.~\ref{Fig: same_cross_coeff_SIMO_more_N} for $N_R \in \{ 8, 16, 32, 64 \}$, with $\rho_{ik}=\rho=0.9$, for $i\neq k$. In Fig.~\ref{Fig: same_cross_coeff_SIMO_N16_32}, the curves for $N_R=4$ are obtained using the following three methods: 1)~based on Theorem~\ref{theo: SIMO_general_N_special_corr}, with the corresponding curves labeled ``Theorem~\ref{theo: SIMO_general_N_special_corr}''; 2)~based on the semi-closed-form expression in Appendix~\ref{App: OrthantP_basics_P4}~\cite{Childs-1967}, with the corresponding curve labeled ``Childs''; and 3)~based on \eqref{eqn: special_h_MMSE_D_I=0} along with a crude version of the algorithm in~\cite{Genz_algorithm} to compute the required orthant probabilities, with the corresponding curve labeled ``Genz''. Since the algorithm in~\cite{Genz_algorithm} is a Monte Carlo method, for the evaluation of each orthant probability, we use $10^4$ times of averaging. The results from all the three methods coincide, which corroborates the analysis leading to Theorem~\ref{theo: SIMO_general_N_special_corr}. However, when $N_R$ becomes large, the semi-closed-form expressions no longer exist, whereas the algorithm in~\cite{Genz_algorithm} becomes computationally more demanding.  
Consequently, we resort to Theorem~\ref{theo: SIMO_general_N_special_corr} to obtain all the curves for the MMSE channel estimate in Fig.~\ref{Fig: same_cross_coeff_SIMO_more_N}. We see in Figs.~\ref{Fig: SIMO_exp_N34_rho59}--\ref{Fig: same_cross_coeff_SIMO_more_N} that, when the channel correlation is high, $\hat{\mathbf{h}}_{\rm MMSE}$ clearly outperforms $\hat{\mathbf{h}}_{\rm BLM}$ across a wide range of SNRs, particularly for moderate-to-high SNRs. The per-antenna MSE of $\hat{\mathbf{h}}_{\rm MMSE}$ improves considerably with $N_R$ for all SNR values. On the other hand, for $\hat{\mathbf{h}}_{\rm BLM}$, the improvement with $N_R$ is predominantly observed for relatively low SNRs.

\subsection{SISO and Spatially White SIMO Channels with Real-Valued Pilot Symbols}

\begin{figure}[t]
\begin{center}
\includegraphics[width=0.99\columnwidth]
{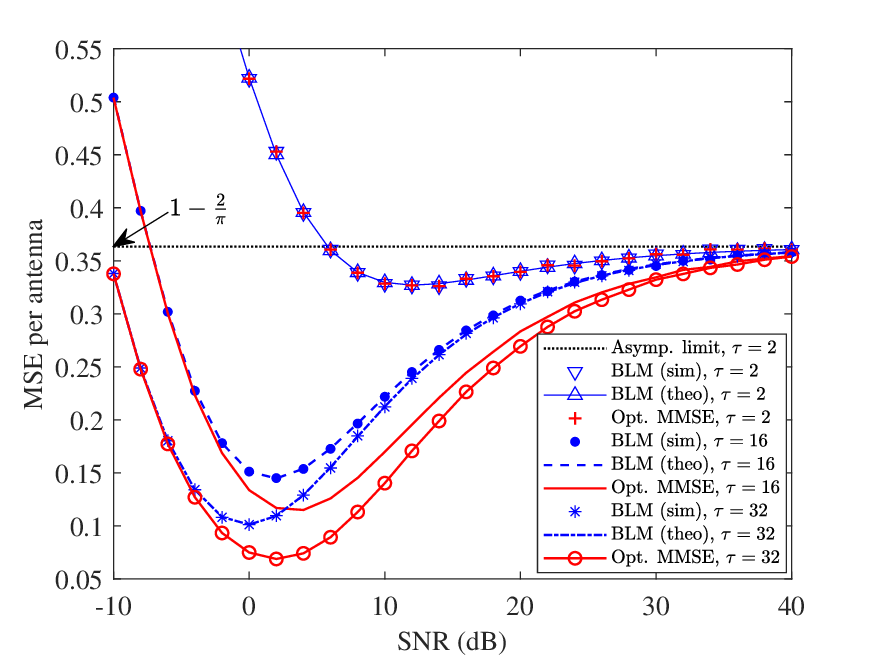}
\end{center}
\caption{MSE comparison for SISO or SIMO with no receive correlation, $\tau \in \{2, 16, 32\}$.}
\label{Fig: multi_pilot_SISO_tau_16_32}
\end{figure}

Here, we investigate the effect of using multiple pilot symbols based on Theorem~\ref{theo: theo_rank_one_SISO}. We use the second column of a $\tau\times\tau$ Hadamard matrix as the pilot vector $\mathbf{s}$. Fig.~\ref{Fig: multi_pilot_SISO_tau_16_32} presents the results for $\tau\in \{ 2, 16, 32\}$. The curves corresponding to $\tau=2$ demonstrate the equivalence of the two estimators as well as the equivalence of \eqref{eqn: equivalence_2_real_pilots} and \eqref{eqn: h_MMSE_SISO_multipilot}. The two estimators exhibit the same asymptotic MSE performance at both low and high SNR. However, considerable improvement in terms of MSE is achievable from nonlinear processing for intermediate SNR values that are of significant practical interest. The numerical results validate the asymptotic MSE for the case with $\tau=2$ mentioned in Remark \ref{remark: tau=2_SISO}. Interestingly, the same behavior at high SNR is also observed for $\tau>2$.  We further observe in Figs.~\ref{Fig: same_cross_coeff_SIMO_N16_32}--\ref{Fig: multi_pilot_SISO_tau_16_32} that, for both $\hat{\mathbf{h}}_{\rm MMSE}$ and $\hat{\mathbf{h}}_{\rm BLM}$, the MSE performance first improves and then deteriorates with the SNR before reaching a limit. This behavior of the MSE performance is a manifestation of the phenomenon referred to as \textit{stochastic resonance}~\cite{J_Mo_R_Heath, Gammaitoni_1995}, where the right amount of noise can be beneficial for the estimation performance in quantized systems. Similar observations can be found in, e.g.,~\cite{Atzeni_2022, Rao_et_al_CH_EST, Utschick_etal_2023}.

\section{Conclusions}\label{sec: conclusions}

In this work, we investigated optimal MMSE channel estimation for MIMO systems in slow, flat Rayleigh fading channels with one-bit quantization at the receiver. We have developed an explicit framework for the computation of the MMSE channel estimate, which is related to the orthant probability of the MVN distribution. Leveraging insights gained from this framework, we determined a condition for the BLMMSE channel estimator to be MSE-optimal. Furthermore, we derived computationally efficient expressions for the nonlinear MMSE channel estimators under specific assumptions regarding channel correlation or pilot symbols. Our findings reveal that the MSE performance of the low-complexity BLMMSE channel estimator closely approximates that of the MMSE estimator under low channel correlation, relatively low SNR, or very high SNR with multiple pilot symbols per transmit antenna. However, with a large number of highly correlated receive antennas or multiple pilot symbols, the MMSE channel estimator yields significant gains over the BLMMSE estimator across a wide range of SNRs. Our results offer unique insights into the one-bit channel estimation problem and provide guidance for applications when $N_R$ or $\tau$ is large.

\begin{appendices}
\section{Orthant Probability in \eqref{eqn: orthant_Prob_formula} with $L=4$}\label{App: OrthantP_basics_P4}

Here, we adopt the same notation as in \eqref{eqn: orthantP_corr_coeff}. Based on~\cite{Childs-1967}, when $L=4$, we have
\begin{align}
\mathcal{P}(\boldsymbol{\Psi}) & = \frac{1}{16} +\frac{\arcsin(\psi_{34}) + \sum_{l=3}^4\arcsin(\psi_{2l})}{8\pi}\nonumber \\
& \phantom{=} \ +\frac{\sum_{k=2}^4 \arcsin(\psi_{1k})}{8\pi}+\frac{\sum_{i=1}^3\mathcal{J}_i}{4\pi^2}, \label{eqn: OrthantP4}
\end{align}
with
\begin{align}
\mathcal{J}_1 & = \displaystyle\int_0^1 \frac{\psi_{12}\arcsin\big(\frac{d_{34, a}-t^2 d_{34, b}}{\sqrt{\left(\mu_{23, a}-t^2\mu_{23, b}\right)\left(\mu_{24, a}-t^2\mu_{24, b}\right)}}\big)}{\sqrt{1-t^2\psi_{12}^2 }}{\rm d}t,
\\\mathcal{J}_2 & = \displaystyle\int_0^1 \frac{\psi_{13}\arcsin\big(\frac{d_{24, a}-t^2 d_{24, b}}{\sqrt{\left(\mu_{23, a}-t^2\mu_{23, b}\right)\left(\mu_{34, a}-t^2\mu_{34, b}\right)}}\big)}{\sqrt{1-t^2\psi_{13}^2}}{\rm d}t,
\\\mathcal{J}_3 & = \displaystyle\int_0^1 \frac{\psi_{14}\arcsin\big(\frac{d_{23, a}-t^2 d_{23, b}}{\sqrt{\left(\mu_{24, a}-t^2\mu_{24, b}\right)\left(\mu_{34, a}-t^2\mu_{34, b}\right)}}\big)}{\sqrt{1-t^2\psi_{14}^2}}{\rm d}t,
\end{align}
and
\begin{align}
d_{23, a} & = \psi_{23}-\psi_{24}\psi_{34}, \\
d_{24, a} & = \psi_{24}-\psi_{23}\psi_{34}, \\
d_{34, a} & = \psi_{34}-\psi_{23}\psi_{24}, \\
\mu_{ik, a} & = 1-\psi_{ik}^2,  \textrm{for}~i<k,~i, k=2,3,4, \\
\mu_{ik, b} & = \psi_{1i}^2+\psi_{1k}^2-2\psi_{1i}\psi_{1k}\psi_{ik},  \textrm{for}~i<k,~i, k=2,3,4, \\
d_{23, b} & = \psi_{12}\psi_{13}+\psi_{14}^2\psi_{23}-\psi_{12}\psi_{14}\psi_{34}-\psi_{13}\psi_{14}\psi_{24}, \\
d_{24, b} & = \psi_{12}\psi_{14}+\psi_{13}^2\psi_{24}-\psi_{12}\psi_{13}\psi_{34}-\psi_{13}\psi_{14}\psi_{23}, \\
d_{34, b} & = \psi_{13}\psi_{14}+\psi_{12}^2\psi_{34}-\psi_{12}\psi_{14}\psi_{23}-\psi_{12}\psi_{13}\psi_{24}.
\end{align}

\section{Derivation of \eqref{eqn: compact_MMSE_form1}}\label{App: main_framework_derivation}
Based on the assumptions in Section~\ref{sec: system_model}, we have
\begin{align}
\hspace{-2mm}\Pr(r_{\mathsf{R}, k}|\mathbf{h}) & = 
\int\limits_0^\infty \frac{\exp\big\{-\frac{1}{\sigma^2}\left(x_k-r_{\mathsf{R}, k}\Re(\mathbf{a}_k^\Trans\mathbf{h})\right)^2\big\}}{\sqrt{\pi\sigma^2}}{\rm d}x_k, \label{new_eqn: framework_fundamental_1}\\
\Pr(r_{\mathsf{I}, k}|\mathbf{h}) & = \int\limits_0^\infty \frac{\exp\big\{-\frac{1}{\sigma^2}\left(y_k-r_{\mathsf{I}, k}\Im(\mathbf{a}_k^\Trans\mathbf{h})\right)^2\big\}}{\sqrt{\pi\sigma^2}}{\rm d}y_k.
\label{new_eqn: framework_fundamental_2}
\end{align}
Since $\Re(n_k)$ and $\Im(n_k)$ are independent, we  further have
\begin{align}
\Pr(r_{k}|\mathbf{h})=\Pr(r_{\mathsf{R}, k}|\mathbf{h})\Pr(r_{\mathsf{I}, k}|\mathbf{h}), \quad \textrm{for}~k=1, \ldots, \tau N_R.\label{eqn: decomp_due_to_noise}
\end{align}
 Due to element-wise quantization, it is clear that
\begin{align}
&\Pr(\mathbf{r}|\mathbf{h}) = \prod_{k=1}^{\tau N_R}\Pr(r_{k}|\mathbf{h}). \label{eqn: decomp_due_to_elementwise_quant} 
\end{align}
 Based on \eqref{new_eqn: framework_fundamental_1}--\eqref{eqn: decomp_due_to_elementwise_quant}
and the assumptions about the noise, we obtain 
\begin{align}
&\Pr(\mathbf{r}|\mathbf{h}) =\displaystyle 
\int_{\mathbb{R}_+^{\tau N_R}}\int_{\mathbb{R}_+^{\tau N_R}}\frac{\exp\{-\mathcal{E}_0\}}{(\pi\sigma^2)^{\tau N_R}}
{\rm d}\mathbf{x} {\rm d}\mathbf{y}, \label{eqn: prob_r_given_h_1}
\end{align}
with 
\begin{align}
\mathcal{E}_0 &=\overset{\tau N_R}{\underset{k=1}{\sum}}\frac{\left(x_k-r_{\mathsf{R}, k}\Re(\mathbf{a}_k^\Trans\mathbf{h})\right)^2+\left(y_k-r_{\mathsf{I}, k}\Im(\mathbf{a}_k^\Trans\mathbf{h})\right)^2}{\sigma^2},
\\
\mathbf{x} &= [x_1 \ldots x_{\tau N_R}]^\Trans, \quad \mathbf{y} = [y_1 \ldots y_{\tau N_R}]^\Trans.
\end{align}
Note that, for $k=1, \ldots, \tau N_R$, we have
\begin{align}\left(\Re(\mathbf{a}_k^\Trans\mathbf{h})\right)^2+\left(\Im(\mathbf{a}_k^\Trans\mathbf{h})\right)^2=\mathbf{h}^\Herm\mathbf{a}_k^*\mathbf{a}_k^\Trans\mathbf{h}.
\end{align}
After some algebra, \eqref{eqn: prob_r_given_h_1} can be expressed as
\begin{align}
\Pr(\mathbf{r}|\mathbf{h})= \int_{\mathbb{R}_+^{\tau N_R}}\int_{\mathbb{R}_+^{\tau N_R}}\frac{\exp\big\{-\frac{\|\boldsymbol{\Lambda}_{\mathsf{R}}\mathbf{x}+j\boldsymbol{\Lambda}_{\mathsf{I}}\mathbf{y} - \mathbf{A}\mathbf{h}\|^2}{\sigma^2}\big\}}{(\pi\sigma^2)^{\tau N_R}}{\rm d}\mathbf{x}{\rm d}\mathbf{y}. \label{eqn: Prob_r_given_h_compact}
\end{align}
Inserting \eqref{eqn: prob_density_h} and \eqref{eqn: Prob_r_given_h_compact} into \eqref{eqn: prob_r_first_appear} yields
\begin{align}
    \Pr(\mathbf{r})= & \int _{\mathbb{R}_+^{\tau N_R}} \! \int_{\mathbb{R}_+^{\tau N_R}} \! \int_{\mathbb{C}^{N_T N_R}} \frac{\exp\left\{-\mathcal{E}_1\right\}}{\pi^{N_T N_R}|\boldsymbol{\Sigma}|(\pi\sigma^2)^{\tau N_R}} {\rm d}\mathbf{h} {\rm d}\mathbf{x} {\rm d}\mathbf{y}, \label{new_eqn: app_1}
\end{align}
with 
\begin{align}
    \mathcal{E}_1 
   &= (\mathbf{h}-\boldsymbol
    {\mu}_{\rm h})^\Herm\boldsymbol{\Theta}^{-1}(\mathbf{h}-\boldsymbol
    {\mu}_{\rm h})+\mathcal{E}_2,  \label{new_eqn: app_2}
      \\\mathcal{E}_2&= \frac{1}{\sigma^2}\|\boldsymbol{\Lambda}_{\mathsf{R}}\mathbf{x}+j\boldsymbol{\Lambda}_{\mathsf{I}}\mathbf{y}\|^2-\boldsymbol
    {\mu}_{\rm h}^\Herm\boldsymbol{\Theta}^{-1}\boldsymbol
    {\mu}_{\rm h}, \label{new_eqn: framework_E2}
\\\boldsymbol{\Theta} &= \left(\boldsymbol{\Sigma}^{-1}+\frac{1}{\sigma^2}\mathbf{A}^\Herm\mathbf{A}\right)^{-1}, 
    \\\boldsymbol{\mu}_{\rm h} &= \frac{1}{\sigma^2}\boldsymbol{\Theta}\mathbf{A}^\Herm\left(\boldsymbol{\Lambda}_{\mathsf{R}}\mathbf{x}+j\boldsymbol{\Lambda}_{\mathsf{I}}\mathbf{y}\right).
    \label{new_eqn: framework_mu_h}
\end{align}
The inner integral in \eqref{new_eqn: app_1} is given by
\begin{align}
   \displaystyle\int_{\mathbb{C}^{N_T N_R}}\exp\{-\mathcal{E}_1\}{\rm d}\mathbf{h} = \pi^{N_T N_R}|\boldsymbol{\Theta}|\exp\{-\mathcal{E}_2\}, \label{new_eqn: app_3}
\end{align}
where we used the fact 
that $\mathcal{CN}(\boldsymbol{\mu}_{\rm h}, \boldsymbol{\Theta})$ integrates to $1$ over $\mathbb{C}^{N_T N_R}$, i.e.,
\begin{align}
     \displaystyle\int_{\mathbb{C}^{N_T N_R}}\frac{\exp\left\{-(\mathbf{h}-\boldsymbol
    {\mu}_{\rm h})^\Herm\boldsymbol{\Theta}^{-1}(\mathbf{h}-\boldsymbol
    {\mu}_{\rm h})\right\}}{ \pi^{N_T N_R}|\boldsymbol{\Theta}|}{\rm d}\mathbf{h}=1.\nonumber
\end{align}
It can be readily verified that \begin{align}
\left|\boldsymbol{\Theta}\right|
    =(\sigma^2)^{\tau N_R}\left|\boldsymbol{\Sigma}\right|/\left|\boldsymbol{\Omega}_{\rm b}\right|.\label{new_eqn: det_of_Theta}
\end{align}
Inserting \eqref{new_eqn: app_3}--\eqref{new_eqn: det_of_Theta} into \eqref{new_eqn: app_1}, after some simplifications, we obtain
\begin{align}
    \Pr(\mathbf{r})= 
&\displaystyle\int _{\mathbb{R}_+^{\tau N_R}}\int_{\mathbb{R}_+^{\tau N_R}} \frac{\exp\left\{-\mathcal{E}_2\right\}}{\pi^{\tau N_R}|\boldsymbol{\Omega}_{\rm b}|} {\rm d}\mathbf{x}{\rm d}\mathbf{y}. \label{new_eqn: app_5}
\end{align}
Using the matrix inversion lemma~\cite{Meyer_matrix}, \eqref{new_eqn: framework_mu_h} is expressed as
\begin{align}
\boldsymbol{\mu}_{\rm h} =\boldsymbol{\Sigma}\mathbf{A}^\Herm\boldsymbol{\Omega}_{\rm b}^{-1}\left(\boldsymbol{\Lambda}_{\mathsf{R}}\mathbf{x}+j\boldsymbol{\Lambda}_{\mathsf{I}}\mathbf{y}\right). 
\label{new_eqn: app_a_added3}
\end{align}
Furthermore, $\mathcal{E}_2$ in \eqref{new_eqn: framework_E2} is equivalent to
\begin{align}
    \mathcal{E}_2 =(\boldsymbol{\Lambda}_{\mathsf{R}}\mathbf{x}+j\boldsymbol{\Lambda}_{\mathsf{I}}\mathbf{y})^\Herm\boldsymbol{\Omega}_{\rm b}^{-1}(\boldsymbol{\Lambda}_{\mathsf{R}}\mathbf{x}+j\boldsymbol{\Lambda}_{\mathsf{I}}\mathbf{y}).\label{new_eqn: app_a_added_2}
\end{align}
Let 
\begin{align}
\mathbf{z} & = \left[\mathbf{x}^\Trans\;\mathbf{y}^\Trans\right]^\Trans \in\mathbb{R}^{2\tau N_R}. \label{eqn: vector_z}
\end{align}
Based on \eqref{eqn: matrix_DR}, \eqref{eqn: matrix_DI}, \eqref{eqn: Matrix_C}, and \eqref{eqn: vector_z}, we can show that \eqref{new_eqn: app_a_added_2} is equivalent to $\mathcal{E}_2=\mathbf{z}^\Trans\mathbf{C}\mathbf{z}$. Inserting $\mathcal{E}_2$ into \eqref{new_eqn: app_5}, we obtain  
\begin{align}
\Pr(\mathbf{r})=
\frac{1}{\pi^{\tau N_R}|\boldsymbol{\Omega}_{\rm b}|}
\int_{\mathbb{R}_+^{2\tau N_R}}
\exp\left\{-\mathbf{z}^\Trans \mathbf{C}\mathbf{z}\right\} {\rm d}\mathbf{z}. \label{eqn:Pr_r_compact}
\end{align}
Similarly, inserting \eqref{eqn: prob_density_h} and \eqref{eqn: Prob_r_given_h_compact} into \eqref{eqn: h_MMSE_txt_formula} yields
\begin{align}
&\displaystyle\int_{\mathbb{C}^{N_T N_R}} \mathbf{h}\Pr(\mathbf{r}|\mathbf{h})p(\mathbf{h}){\rm d}\mathbf{h} \nonumber
     \\&=  
    \int _{\mathbb{R}_+^{\tau N_R}}\int_{\mathbb{R}_+^{\tau N_R}} \int_{\mathbb{C}^{N_T N_R}} \frac{\mathbf{h}\exp\left\{-\mathcal{E}_1\right\}}{\pi^{N_T N_R}|\boldsymbol{\Sigma}|(\pi\sigma^2)^{\tau N_R}} {\rm d}\mathbf{h} {\rm d}\mathbf{x}{\rm d}\mathbf{y}. \label{new_eqn: app_num_1}
\end{align}
By the definition of the mean vector of $\mathcal{CN}(\boldsymbol{\mu}_{\rm h}, \boldsymbol{\Theta})$, we have
\begin{align*}
 &\displaystyle\int_{\mathbb{C}^{N_T N_R}}\frac{\mathbf{h}\exp\left\{-(\mathbf{h}-\boldsymbol
    {\mu}_{\rm h})^\Herm\boldsymbol{\Theta}^{-1}(\mathbf{h}-\boldsymbol
    {\mu}_{\rm h})\right\}}{ \pi^{N_T N_R}|\boldsymbol{\Theta}|} {\rm d}\mathbf{h}=\boldsymbol{\mu}_{\rm h}.
    \end{align*}
Then, the inner integral in \eqref{new_eqn: app_num_1} is given by
\begin{align}
\displaystyle\int_{\mathbb{C}^{N_T N_R}}\mathbf{h}\exp\left\{-\mathcal{E}_1\right\}{\rm d}\mathbf{h} = \pi^{N_T N_R}|\boldsymbol{\Theta}|\exp\{-\mathcal{E}_2\}\boldsymbol{\mu}_{\rm h}.\label{new_eqn: app_num_2}
\end{align}
Now, we insert \eqref{new_eqn: app_num_2} into \eqref{new_eqn: app_num_1}. Using \eqref{new_eqn: det_of_Theta}, \eqref{new_eqn: app_a_added3} and the fact that $\boldsymbol{\Lambda}_{\mathsf{R}}\mathbf{x}+j\boldsymbol{\Lambda}_{\mathsf{I}}\mathbf{y}=\left[\boldsymbol{\Lambda}_{\mathsf{R}} \ j\boldsymbol{\Lambda}_{\mathsf{I}}\right]\mathbf{z}$, we follow similar steps as for the development of \eqref{eqn:Pr_r_compact} to obtain
\begin{align}
&\displaystyle\int_{\mathbb{C}^{N_T N_R}} \mathbf{h}\Pr(\mathbf{r}|\mathbf{h})p(\mathbf{h}){\rm d}\mathbf{h} \nonumber
\\& = \frac{\boldsymbol{\Sigma}\mathbf{A}^\Herm\boldsymbol{\Omega}_{\rm b}^{-1}\left[\boldsymbol{\Lambda}_{\mathsf{R}} \ j\boldsymbol{\Lambda}_{\mathsf{I}}\right]}{\pi^{\tau N_R}|\boldsymbol{\Omega}_{\rm b}|} \int_{\mathbb{R}_+^{2\tau N_R}}
\mathbf{z}\exp\left\{-\mathbf{z}^\Trans \mathbf{C}\mathbf{z}\right\}{\rm d}\mathbf{z}.
\label{eqn: h_MMSE_numerator_final_compact}
\end{align}
Finally, with \eqref{eqn:Pr_r_compact} and \eqref{eqn: h_MMSE_numerator_final_compact}, we express  \eqref{eqn: h_MMSE_txt_formula} as in \eqref{eqn: compact_MMSE_form1}.

\section{Derivation of \eqref{eqn: reduction_2} from \eqref{eqn: compact_MMSE_form1}}\label{App: deriving_h_MMSE_using_Orthant_P}

Using the Gaussian orthant probability in \eqref{eqn: orthant_Prob_formula} and the property in \eqref{eqn: OrthantP_scaling}, we obtain
 \begin{align}
\int_{\mathbb{R}_+^{2\tau N_R}} 
\exp\{-\mathbf{z}^\Trans \mathbf{C}\mathbf{z}\} {\rm d}\mathbf{z} &=(2\pi)^{\tau N_R}\left|\frac{1}{2}\mathbf{C}^{-1}\right|^{\frac{1}{2}}\mathcal{P}\left(\frac{1}{2}\mathbf{C}^{-1}\right)\nonumber
\\&=\pi^{\tau N_R}\left|\mathbf{C}\right|^{-\frac{1}{2}}\mathcal{P}\left(\mathbf{C}^{-1}\right).
\label{eqn: h_MMSE_orthantP_tmp1}
\end{align}
Let $\mathbf{z}_{-k} \in\mathbb{R}^{2\tau N_R-1}$ denote the vector obtained by deleting the $k$-th element of $\mathbf{z}$,
and $\mathbf{C}_{-k}\in\mathbb{R}^{(2\tau N_R-1)\times (2\tau N_R-1)}$ the matrix obtained by deleting the $k$-th row and column of $\mathbf{C}$, for $k=1, \ldots, 2\tau N_R$. Using vector calculus, we have 
\begin{align}
\nabla_{\mathbf{z}}\exp\left\{-\mathbf{z}^\Trans \mathbf{C}\mathbf{z}\right\}
&=\left[\frac{\partial\exp\left\{-\mathbf{z}^\Trans\mathbf{C}\mathbf{z}\right\}}{\partial z_k}\right]_{k=1}^{2\tau N_R} \label{new_eqn: grad_1}
\\
&=-2\mathbf{Cz} \exp\left\{-\mathbf{z}^\Trans \mathbf{C}\mathbf{z}\right\} \label{new_eqn: grad_2}
\end{align}
and
\begin{align}
&\int_{\mathbb{R}_+^{2\tau N_R}} \mathbf{z}
    \exp\left\{-\mathbf{z}^\Trans \mathbf{C}\mathbf{z}\right\} {\rm d}\mathbf{z} \nonumber
    \\& = -\frac{\mathbf{C}^{-1}}{2}\int_{\mathbb{R}_+^{2\tau N_R}}\nabla_{\mathbf{z}}\exp\left\{-\mathbf{z}^\Trans \mathbf{C}\mathbf{z}\right\} {\rm d}\mathbf{z} 
    \label{new_eqn: grad_app_1}
     \\& =-\frac{\mathbf{C}^{-1}}{2}
   \left[
\displaystyle\int_{\mathbb{R}_+^{2\tau N_R-1}} \displaystyle\int_0^\infty\frac{\partial\exp\left\{-\mathbf{z}^\Trans\mathbf{C}\mathbf{z}\right\}}{\partial z_k}{\rm d}z_k {\rm d}\mathbf{z}_{-k} 
   \right]_{k=1}^{2\tau N_R},\label{new_eqn: grad_app_2}
   \end{align}
where in~\eqref{new_eqn: grad_app_1}  we applied~\eqref{new_eqn: grad_2} and in~\eqref{new_eqn: grad_app_2}  we applied~\eqref{new_eqn: grad_1}. The inner integral in the $k$-th element of~\eqref{new_eqn: grad_app_2} is given by
\begin{align}
\displaystyle\int_0^\infty\frac{\partial\exp\left\{-\mathbf{z}^\Trans\mathbf{C}\mathbf{z}\right\}}{\partial z_k}{\rm d}z_k = \left.\exp\left\{-\mathbf{z}^\Trans\mathbf{C}\mathbf{z}\right\}\right|_{z_k=0}^\infty. \label{new_eqn: grad_app_3}
\end{align}
Clearly, we have
\begin{align}
\lim\limits_{z_k\rightarrow\infty}\exp\left\{-\mathbf{z}^{\mathrm{T}}\mathbf{C}\mathbf{z}\right\}=0.
\end{align}
Further defining
\begin{align}
\breve{\mathbf{z}}_{k}=[z_1\ldots z_{k-1} \; 0\; z_{k+1} \ldots z_{2\tau N_R}]^\Trans,
\end{align}
we have
    \begin{align}
      \left. \mathbf{z}^{\mathrm{T}}\mathbf{C}\mathbf{z}\right|_{z_k=0}=\breve{\mathbf{z}}_{k}^\Trans\mathbf{C}\breve{\mathbf{z}}_{k}= \mathbf{z}_{-k}^{\mathrm{T}}\mathbf{C}_{-k}\mathbf{z}_{-k}.    \end{align}
 Therefore, from~\eqref{new_eqn: grad_app_3}, we obtain
 \begin{align}
\displaystyle\int_0^\infty\frac{\partial\exp\left\{-\mathbf{z}^\Trans\mathbf{C}\mathbf{z}\right\}}{\partial z_k}{\rm d}z_k = -\exp\left\{-\mathbf{z}_{-k}^{\mathrm{T}}\mathbf{C}_{-k}\mathbf{z}_{-k}\right\}. \label{new_eqn: grad_app_4}
\end{align}
Inserting~\eqref{new_eqn: grad_app_4} into~\eqref{new_eqn: grad_app_2} yields
   \begin{align}
&\int_{\mathbb{R}_+^{2\tau N_R}} \mathbf{z}
    \exp\left\{-\mathbf{z}^\Trans \mathbf{C}\mathbf{z}\right\} {\rm d}\mathbf{z} \nonumber
        \\& =\frac{\mathbf{C}^{-1}}{2}\left[\int_{\mathbb{R}_+^{2\tau N_R-1}} \exp\left\{-\mathbf{z}_{-k}^\Trans\mathbf{C}_{-k}\mathbf{z}_{-k}\right\}{\rm d}\mathbf{z}_{-k} \right]_{k=1}^{2\tau N_R}
    \\& =\frac{\mathbf{C}^{-1}}{2}\left[\pi^{\frac{2\tau N_R-1}{2}}\left|\mathbf{C}_{-k}\right|^{-\frac{1}{2}}\mathcal{P}\left((\mathbf{C}_{-k})^{-1}\right)  \right]_{k=1}^{2\tau N_R} \label{eqn: alpha_k_tmp}
    \\& =\frac{1}{2}\pi^{\frac{2\tau N_R-1}{2}}\mathbf{C}^{-1}\Diag{}\left(\left[\big|\mathbf{C}_{-k}\right|^{-\frac{1}{2}}\big]_{k=1}^{2\tau N_R}\right)\mathbf{g}, \label{eqn: method_reduction_2}
\end{align}
where the $k$-th element of $\mathbf{g}$ is given in \eqref{eqn: new_g_k}
and where in \eqref{eqn: alpha_k_tmp} we used \eqref{eqn: OrthantP_scaling}.
It can also be shown that
\begin{align}
|\mathbf{C}_{-k}|/|\mathbf{C}|=[\mathbf{C}^{-1}]_{kk}, \quad \textrm{for}~k=1, \ldots, 2\tau N_R. \label{eqn: h_MMSE_orthantP_tmp2_det}
\end{align} 
Inserting \eqref{eqn: h_MMSE_orthantP_tmp1} and \eqref{eqn: method_reduction_2}  into \eqref{eqn: compact_MMSE_form1}, using \eqref{eqn: h_MMSE_orthantP_tmp2_det}, and after some algebra, we obtain
\begin{align}
    \frac{\displaystyle \int_{\mathbb{R}_+^{2\tau N_R}} \mathbf{z} \exp\{-\mathbf{z}^\Trans\mathbf{C}\mathbf{z}\} {\rm d}\mathbf{z}}{\displaystyle\int_{\mathbb{R}_+^{2\tau N_R}} \exp\{-{\mathbf{z}}^\Trans \mathbf{C}{\mathbf{z}}\} {\rm d}{\mathbf{z}}}=\frac{ \mathbf{C}^{-1}\big(\widetilde{\Diag{}}(\mathbf{C}^{-1})\big)^{-\frac{1}{2}}\mathbf{g}}
{2\sqrt{\pi}\mathcal{P}\left(\mathbf{C}^{-1}\right)}. \label{eqn: useful_quotient_mean_vec}
\end{align}
Finally, \eqref{eqn: reduction_2} follows from inserting \eqref{eqn: useful_quotient_mean_vec} into \eqref{eqn: compact_MMSE_form1}.

\section{Proof of Lemma~\ref{lemma: OrthantP_2_linearity}}\label{App_proof_Lemma}

From \eqref{eqn: useful_quotient_mean_vec} and \eqref{eqn: OrthantP_scaling}, it is clear that
\begin{align}
\eqref{eqn: key_expression_in_lemma1_for_theorem}=\frac{\boldsymbol{\Lambda}_\mathbf{q}\boldsymbol{\Phi}^{-1}\big(\widetilde{\Diag{}}(\boldsymbol{\Phi}^{-1})\big)^{-\frac{1}{2}}\left[\mathcal{P}\left((\boldsymbol{\Phi}_{-i})^{-1}\right)\right]_{i=1}^2}{2\sqrt{\pi}\mathcal{P}(\boldsymbol{\Phi}^{-1})},
\label{eqn: lemma1_for_theorem1_App1}
\end{align} 
with $\mathcal{P}\left((\boldsymbol{\Phi}_{-i})^{-1}\right)=\frac{1}{2}$, for $i=1, 2$. Using \eqref{eqn: OrthantP_identity} and \eqref{eqn: OrthantP_N2} to obtain 
\begin{align}
\mathcal{P}(\boldsymbol{\Phi}^{-1})=\frac{1}{4}-\frac{q_1q_2\arcsin(\phi_{12})}{2\pi},
\end{align}
and capitalizing on the fact that $q_1, q_2\in\{\pm 1\}$, after some algebra, we obtain
\begin{align}
\eqref{eqn: key_expression_in_lemma1_for_theorem} = \frac{\begin{bmatrix}
\breve{\sigma}_{2}\big(1-\frac{\phi_{12}\arcsin(\phi_{12})}{\pi/2}\big) & \breve{\sigma}_{2}\big(\frac{\arcsin(\phi_{12})}{\pi/2}-\phi_{12}\big) \\
\breve{\sigma}_{1}\big(\frac{\arcsin(\phi_{12})}{\pi/2}-\phi_{12}\big) & \breve{\sigma}_{1}\big(1-\frac{\phi_{12}\arcsin(\phi_{12})}{\pi/2}\big)
\end{bmatrix}\mathbf{q}}{\breve{\sigma}_{1}\breve{\sigma}_{2}\sqrt{\pi(1-\phi_{12}^2)}\big(1-\big(\frac{\arcsin(\phi_{12})}{\pi/2}\big)^2\big)},\label{eqn: lemma_linearity_N2_main}
\end{align}
which is linear in $\mathbf{q}$. \hfill \IEEEQED

\section{Proof of Lemma~\ref{lemma: OrthantP_3_nonL}}\label{app: loss_of_linearity_N3}

With a slight abuse of notation, we use a notation similar to that in Appendix~\ref{App_proof_Lemma}.  Based on \eqref{eqn: useful_quotient_mean_vec}, we have
\begin{align}
\eqref{eqn: key_expression_in_lemma2_for_theorem} = \frac{\boldsymbol{\Lambda}_\mathbf{q}\boldsymbol{\Phi}^{-1}\big(\widetilde{\Diag{}}(\boldsymbol{\Phi}^{-1})\big)^{-\frac{1}{2}}\left[\mathcal{P}\left((\boldsymbol{\Phi}_{-i})^{-1}\right)\right]_{i=1}^3}{2\sqrt{\pi}\mathcal{P}(\boldsymbol{\Phi}^{-1})}.
\label{eqn: lemma2_for_theorem1_App2}
 \end{align}
 From \eqref{eqn: OrthantP_N3}, we readily obtain
\begin{align}
\mathcal{P}(\boldsymbol{\Phi}^{-1})&=\frac{1}{8}+\frac{q_1q_2\alpha_{12}+q_1q_3\alpha_{13}+q_2q_3\alpha_{23}}{4\pi}, \label{eqn: lemma2_proof_1}
\end{align}
with
\begin{align}
\alpha_{ik}=\arcsin\left(\frac{\phi_{il}\phi_{kl}-\phi_{ik}}{\sqrt{1-\phi_{il}^2}\sqrt{1-\phi_{kl}^2}}\right), 
\end{align} 
for $i, k, l=1, 2, 3$, $i\neq k$, $i\neq l$, and $k\neq l$. From \eqref{eqn: OrthantP_N2}, we can further show that
\begin{align}
\mathcal{P}\left((\boldsymbol{\Phi}_{-i})^{-1}\right)=\frac{1}{4}-\frac{q_kq_l\arcsin(\phi_{kl})}{2\pi},
\end{align}
for $i, k, l=1, 2, 3$, $i\neq k$, $i\neq l$, and $k\neq l$. Using the fact that $q_i\in\{\pm 1\},$ for $i=1, 2, 3$, we can conclude that
\begin{align}
&
\eqref{eqn: key_expression_in_lemma2_for_theorem} = \frac{\mathbf{T}_{\rm lem}}{2\sqrt{\pi c_{\rm lem}}\;\mathcal{P}(\boldsymbol{\Phi}^{-1})}\begin{bmatrix}
    \frac{q_1}{4}-\frac{q_1q_2q_3\arcsin(\phi_{23})}{2\pi}
    \\ \frac{q_2}{4}-\frac{q_1q_2q_3\arcsin(\phi_{13})}{2\pi}
    \\ \frac{q_3}{4}-\frac{q_1q_2q_3\arcsin(\phi_{12})}{2\pi}
\end{bmatrix}, \label{new_eqn: lem2_3by3_added}
\end{align}
with
\begin{align}
c_{\rm lem}&=1+2\phi_{12}\phi_{23}\phi_{13}-(\phi_{12}^2+\phi_{23}^2+\phi_{13}^3),
    \\\mathbf{T}_{\rm lem}&=\begin{bmatrix}
\frac{\sqrt{1-\phi_{23}^2}}{\breve{\sigma}_{1}} & \frac{\phi_{13}\phi_{23}-\phi_{12}}{\breve{\sigma}_{1}\sqrt{1-\phi_{13}^2}} &\frac{\phi_{12}\phi_{23}-\phi_{13}}{\breve{\sigma}_{1}\sqrt{1-\phi_{12}^2}} \\
\frac{\phi_{13}\phi_{23}-\phi_{12}}{\breve{\sigma}_{2}\sqrt{1-\phi_{23}^2}}  & \frac{\sqrt{1-\phi_{13}^2}}{\breve{\sigma}_{2}} &\frac{\phi_{12}\phi_{13}-\phi_{23}}{\breve{\sigma}_{2}\sqrt{1-\phi_{12}^2}}\\
\frac{\phi_{12}\phi_{23}-\phi_{13}}{\breve{\sigma}_{3}\sqrt{1-\phi_{23}^2}} &\frac{\phi_{12}\phi_{13}-\phi_{23}}{\breve{\sigma}_{3}\sqrt{1-\phi_{13}^2}} & \frac{\sqrt{1-\phi_{12}^2}}{\breve{\sigma}_{3}}
\end{bmatrix}.
\end{align}
Through detailed calculations, we can observe that \eqref{eqn: key_expression_in_lemma2_for_theorem} (or \eqref{new_eqn: lem2_3by3_added}) is nonlinear in $\mathbf{q}$ when $\boldsymbol{\Phi}$ is a full matrix with all non-zero elements or when only one correlation coefficient  in $\boldsymbol{\Phi}$ is zero (i.e., $\phi_{12}=0$, $\phi_{13}=0$, or $\phi_{23}=0$). \hfill \IEEEQED

\section{Proof of Theorem~\ref{theo: Theorem_BLMMSE_opt}}\label{App: proof_Theorem_BLMMSE_opt}

Suppose that each row of $\mathbf{C}$ contains at most two non-zero elements. Then, there exists a suitable permutation matrix $\boldsymbol{\Gamma}$ satisfying
\begin{align}
\boldsymbol{\Gamma}\boldsymbol{\Gamma}^\Trans&=\boldsymbol{\Gamma}^\Trans\boldsymbol{\Gamma}=\mathbf{I}_{2\tau N_R} \label{eqn: perm_matrix_proof_theorem1}
\end{align}
such that $\boldsymbol{\Gamma}\mathbf{C}\boldsymbol{\Gamma}^\Trans$ is block-diagonal with exactly $\tau N_R$ blocks, each of size $2\times 2$, along its diagonal. Let 
\begin{align}
\boldsymbol{\Lambda}_{\mathsf{C}}&=\boldsymbol{\Gamma}\BDiag(\boldsymbol{\Lambda}_{\mathsf{R}}, \boldsymbol{\Lambda}_{\mathsf{I}}) 
\boldsymbol{\Gamma}^\Trans, \label{new_eqn: Theo1_proof_added}
\end{align}
which is clearly diagonal (cf. \eqref{eqn:quantized_output_diag_mat}), and let
\begin{align}\mathbf{M}_{\mathsf{C}}&=\boldsymbol{\Gamma}\begin{bmatrix}
    \mathbf{D}_{\mathsf{R}} &\mathbf{D}_{\mathsf{I}}^\Trans\\\mathbf{D}_{\mathsf{I}}& \mathbf{D}_{\mathsf{R}}
\end{bmatrix}\boldsymbol{\Gamma}^\Trans.
\end{align}
From \eqref{eqn: Matrix_C} and \eqref{eqn: perm_matrix_proof_theorem1}, it is straightforward to see that  $\boldsymbol{\Gamma}\mathbf{C}\boldsymbol{\Gamma}^\Trans =\boldsymbol{\Lambda}_{\mathsf{C}}\mathbf{M}_{\mathsf{C}}\boldsymbol{\Lambda}_{\mathsf{C}}$ and, thus, $\mathbf{M}_{\mathsf{C}}$ must be block-diagonal, also with $2\times 2$ blocks along its diagonal. Let us write
\begin{align}
\mathbf{M}_{\mathsf{C}}=\BDiag(\mathbf{M}_1, \ldots,  \mathbf{M}_{\tau N_R}) \label{eqn: def_matrix_M_C}
\end{align}
and consider \eqref{eqn: compact_MMSE_form1}. Applying the change of variables $\mathbf{t}=\boldsymbol{\Gamma}\mathbf{z}$ in the integrals of \eqref{eqn: compact_MMSE_form1}, using \eqref{eqn: perm_matrix_proof_theorem1}, we obtain
\begin{align}
\hat{\mathbf{h}}_{\rm MMSE} & = \boldsymbol{\Sigma}\mathbf{A}^\Herm\boldsymbol{\Omega}_{\rm b}^{-1}\left[\mathbf{I}_{\tau N_R} \ j\mathbf{I}_{\tau N_R}\right]\boldsymbol{\Gamma}^\Trans \nonumber\\
& \phantom{=} \ \cdot \frac{\boldsymbol{\Lambda}_{\mathsf{C}}\displaystyle 
\int_{\mathbb{R}_+^{2\tau N_R}}
\mathbf{t} \exp\left\{-\mathbf{t}^\Trans\boldsymbol{\Lambda}_{\mathsf{C}}\mathbf{M}_{\mathsf{C}}\boldsymbol{\Lambda}_{\mathsf{C}} \mathbf{t}\right\} {\rm d}\mathbf{t}}
{\displaystyle 
\int_{\mathbb{R}_+^{2\tau N_R}}
\exp\left\{-\mathbf{t}^\Trans\boldsymbol{\Lambda}_{\mathsf{C}}\mathbf{M}_{\mathsf{C}}\boldsymbol{\Lambda}_{\mathsf{C}}\mathbf{t}\right\} {\rm d}\mathbf{t}}.
\label{eqn: Theorem_linearity_tmp1}
\end{align} 
At this stage, we can observe the striking similarity between the quotient in \eqref{eqn: Theorem_linearity_tmp1} and \eqref{eqn: key_expression_in_lemma1_for_theorem}, as well as between \eqref{eqn: Matrix_Phi_in_lemma1} and  
the matrix $\boldsymbol{\Lambda}_{\mathsf{C}}\mathbf{M}_{\mathsf{C}}\boldsymbol{\Lambda}_{\mathsf{C}}$. Clearly, what we need is to break up the integral in the denominator of the quotient in \eqref{eqn: Theorem_linearity_tmp1} into a product of integrals over $\mathbb{R}_+^{2}$. To this end, let 
\begin{align}
\mathbf{t} &=[\widetilde{\mathbf{t}}_i]_{i=1}^{\tau N_R}=[\widetilde{\mathbf{t}}_1^\Trans \ldots \widetilde{\mathbf{t}}_{\tau N_R}^\Trans]^\Trans ,\label{eqn: specific_vec_notation_used_later}
\\ \boldsymbol{\Lambda}_{\mathsf{C}}&=\BDiag(\boldsymbol{\Lambda}_{\mathsf{Cblk}, 1}, \ldots, \boldsymbol{\Lambda}_{\mathsf{Cblk}, \tau N_R}),\label{eqn: specific_mat_notation_used_later}
\end{align}
where $\widetilde{\mathbf{t}}_i$ consists of $t_{2i-1}$ and $t_{2i}$ from $\mathbf{t}$, and the diagonal elements of $\boldsymbol{\Lambda}_{\mathsf{Cblk}, i}$ are the $(2i-1, 2i-1)$-th and $(2i, 2i)$-th elements of  $\boldsymbol{\Lambda}_{\mathsf{C}}$, for $i=1, \ldots, \tau N_R$. 
Using \eqref{eqn: def_matrix_M_C} and \eqref{eqn: specific_vec_notation_used_later}--\eqref{eqn: specific_mat_notation_used_later}, we have 
\begin{align}
    &\displaystyle 
\int_{\mathbb{R}_+^{2\tau N_R}}
\exp\left\{-\mathbf{t}^\Trans\boldsymbol{\Lambda}_{\mathsf{C}}\mathbf{M}_{\mathsf{C}}\boldsymbol{\Lambda}_{\mathsf{C}}\mathbf{t}\right\} {\rm d}\mathbf{t}\nonumber
\\&=\prod_{i=1}^{\tau N_R}\int_{\mathbb{R}_+^{2}}
\exp\left\{-\widetilde{\mathbf{t}}_i^\Trans\boldsymbol{\Lambda}_{\mathsf{Cblk}, i} \mathbf{M}_i\boldsymbol{\Lambda}_{\mathsf{Cblk}, i} \widetilde{\mathbf{t}}_i\right\} {\rm d}\widetilde{\mathbf{t}}_i
\label{eqn: Theo1_proof_big1}
\end{align}
and, correspondingly, 
\begin{align}
&\boldsymbol{\Lambda}_{\mathsf{C}}\displaystyle 
\int_{\mathbb{R}_+^{2\tau N_R}}\mathbf{t}
\exp\{-\mathbf{t}^\Trans\boldsymbol{\Lambda}_{\mathsf{C}}\mathbf{M}_{\mathsf{C}}\boldsymbol{\Lambda}_{\mathsf{C}}\mathbf{t}\} {\rm d}\mathbf{t}\nonumber
\\&=\Bigg[\boldsymbol{\Lambda}_{\mathsf{Cblk}, k}\int_{\mathbb{R}_+^{2}}\widetilde{\mathbf{t}}_k
\exp\{-\widetilde{\mathbf{t}}_k^\Trans\boldsymbol{\Lambda}_{\mathsf{Cblk}, k} \mathbf{M}_k\boldsymbol{\Lambda}_{\mathsf{Cblk}, k} \widetilde{\mathbf{t}}_k\} {\rm d}\widetilde{\mathbf{t}}_k \nonumber
\\&\phantom{=} \ \cdot\Bigg(\prod_{\substack{i=1\\i\neq k}}^{\tau N_R}\int_{\mathbb{R}_+^{2}}
\exp\{-\widetilde{\mathbf{t}}_i^\Trans\boldsymbol{\Lambda}_{\mathsf{Cblk}, i} \mathbf{M}_i\boldsymbol{\Lambda}_{\mathsf{Cblk}, i} \widetilde{\mathbf{t}}_i\} {\rm d}\widetilde{\mathbf{t}}_i\Bigg)\Bigg]_{k=1}^{\tau N_R}.\label{eqn: Theo1_proof_big2}
\end{align}
Inserting \eqref{eqn: Theo1_proof_big1}--\eqref{eqn: Theo1_proof_big2} into \eqref{eqn: Theorem_linearity_tmp1}, after some algebra, we obtain
\begin{align}
&\hat{\mathbf{h}}_{\rm MMSE} = \boldsymbol{\Sigma}\mathbf{A}^\Herm\boldsymbol{\Omega}_{\rm b}^{-1}\left[\mathbf{I}_{\tau N_R} \ j\mathbf{I}_{\tau N_R}\right]\boldsymbol{\Gamma}^\Trans \nonumber\\
& \cdot \left[\frac{\boldsymbol{\Lambda}_{\mathsf{Cblk}, k}\displaystyle\int_{\mathbb{R}_+^{2}}\widetilde{\mathbf{t}}_k
\exp\left\{-\widetilde{\mathbf{t}}_k^\Trans\boldsymbol{\Lambda}_{\mathsf{Cblk}, k} \mathbf{M}_k\boldsymbol{\Lambda}_{\mathsf{Cblk}, k} \widetilde{\mathbf{t}}_k\right\} {\rm d}\widetilde{\mathbf{t}}_k}
{\displaystyle\int_{\mathbb{R}_+^{2}}
\exp\left\{-\widetilde{\mathbf{t}}_k^\Trans\boldsymbol{\Lambda}_{\mathsf{Cblk}, k} \mathbf{M}_k\boldsymbol{\Lambda}_{\mathsf{Cblk}, k} \widetilde{\mathbf{t}}_k\right\} {\rm d}\widetilde{\mathbf{t}}_k}\right]_{k=1}^{\tau N_R}
\label{eqn: Theorem_linearity_tmp2}
\end{align} 
as the final result. 
Each block element of \eqref{eqn: Theorem_linearity_tmp2} has a structure identical to \eqref{eqn: key_expression_in_lemma1_for_theorem}. Based on Lemma~\ref{lemma: OrthantP_2_linearity}, we conclude  that
$\hat{\mathbf{h}}_{\rm MMSE}$ is linear in the diagonal elements of $\boldsymbol{\Lambda}_{\mathsf{C}}$ in \eqref{new_eqn: Theo1_proof_added} and, hence, it is linear in  $\mathbf{r}$. Since $\hat{\mathbf{h}}_{\rm BLM}$ yields the smallest MSE among the class of linear estimators~\cite{Y_Li_et_al_BLMMSE, S-M-Kay_estimation}, it must be MSE-optimal and equivalent to $\hat{\mathbf{h}}_{\rm MMSE}$ in this case. On the other hand, if there are more than two non-zero elements in one row of $\mathbf{C}$, the covariance-type matrix $\mathbf{C}$ indicates correlation among three or more variables. Then, there exists at least one sub-matrix of $\mathbf{C}$ corresponding to this correlation among three variables.  Therefore, based on Lemma~\ref{lemma: OrthantP_3_nonL}, $\hat{\mathbf{h}}_{\rm MMSE}$ cannot be linear in $\mathbf{r}$ in this case. \hfill \IEEEQED

\section{A Special Case of the MMSE Channel Estimate} \label{App: special_case_h_MMSE}

Consider the special case where $\boldsymbol{\Omega}_{\rm b}$ in \eqref{eqn: corr_mat_of_b} is real-valued and $\boldsymbol{\Lambda}_{\mathsf{I}}=\mathbf{0}$. Clearly, we have
\begin{align}
\mathbf{C}&=\BDiag{(\boldsymbol{\Lambda}_{\mathsf{R}}\mathbf{D}_{\mathsf{R}}\boldsymbol{\Lambda}_{\mathsf{R}}, \boldsymbol{\Lambda}_{\mathsf{I}}\mathbf{D}_{\mathsf{R}}\boldsymbol{\Lambda}_{\mathsf{I}})},
\\\mathbf{C}^{-1}&=\BDiag{(\boldsymbol{\Lambda}_{\mathsf{R}}\boldsymbol{\Omega}_{\rm b}\boldsymbol{\Lambda}_{\mathsf{R}}, \boldsymbol{\Lambda}_{\mathsf{I}}\boldsymbol{\Omega}_{\rm b}\boldsymbol{\Lambda}_{\mathsf{I}})}.
\end{align}
Through straightforward calculations, we obtain
\begin{align}
    \mathcal{P}(\mathbf{C}^{-1})=\mathcal{P}(\boldsymbol{\Lambda}_{\mathsf{R}}\boldsymbol{\Omega}_{\rm b}\boldsymbol{\Lambda}_{\mathsf{R}})\cdot\mathcal{P}(\boldsymbol{\Lambda}_{\mathsf{I}}\boldsymbol{\Omega}_{\rm b}\boldsymbol{\Lambda}_{\mathsf{I}}). \label{eqn: spec_1}
\end{align}
Due to the block-diagonal structure of $\mathbf{C}$, we have 
\begin{align}
    \mathcal{P}\left((\mathbf{C}_{-k})^{-1}\right)& =\mathcal{P}(((\boldsymbol{\Lambda}_{\mathsf{R}}\mathbf{D}_{\mathsf{R}}\boldsymbol{\Lambda}_{\mathsf{R}})_{-k})^{-1})\mathcal{P}(\boldsymbol{\Lambda}_{\mathsf{I}}\boldsymbol{\Omega}_{\rm b}\boldsymbol{\Lambda}_{\mathsf{I}}), \label{eqn: spec_2} \\
\mathcal{P}\left((\mathbf{C}_{-(k+\tau N_R)})^{-1}\right) &=
\mathcal{P}(\boldsymbol{\Lambda}_{\mathsf{R}}\boldsymbol{\Omega}_{\rm b}\boldsymbol{\Lambda}_{\mathsf{R}})\mathcal{P}\big(\big((\boldsymbol{\Lambda}_{\mathsf{I}}\mathbf{D}_{\mathsf{R}}\boldsymbol{\Lambda}_{\mathsf{I}})_{-k}\big)^{-1}\big), \label{eqn: spec_3}
\end{align}
for $k=1, \ldots, \tau N_R$. Applying \eqref{eqn: spec_1}--\eqref{eqn: spec_3} to \eqref{eqn: reduction_2}, we obtain 
\begin{align}
\hat{\mathbf{h}}_{\rm MMSE} & = \frac{\boldsymbol{\Sigma}\mathbf{A}^\Herm\mathbf{D}_{\Omega}^{-\frac{1}{2}}}
{2\sqrt{\pi}}\left(\frac{\boldsymbol{\Lambda}_{\mathsf{R}}\big[\mathcal{P}\big(\big((\boldsymbol{\Lambda}_{\mathsf{R}}\mathbf{D}_{\mathsf{R}}\boldsymbol{\Lambda}_{\mathsf{R}})_{-k}\big)^{-1}\big)\big]_{k=1}^{\tau N_R}}{\mathcal{P}(\boldsymbol{\Lambda}_{\mathsf{R}}\boldsymbol{\Omega}_{\rm b}\boldsymbol{\Lambda}_{\mathsf{R}})}\right. \nonumber \\
   & \phantom{=} \ \left.+\frac{j\boldsymbol{\Lambda}_{\mathsf{I}}\big[\mathcal{P}\big(\big((\boldsymbol{\Lambda}_{\mathsf{I}}\mathbf{D}_{\mathsf{R}}\boldsymbol{\Lambda}_{\mathsf{I}})_{-k}\big)^{-1}\big)\big]_{k=1}^{\tau N_R}}{\mathcal{P}(\boldsymbol{\Lambda}_{\mathsf{I}}\boldsymbol{\Omega}_{\rm b}\boldsymbol{\Lambda}_{\mathsf{I}})}\right), \label{eqn: special_h_MMSE_D_I=0}
   \end{align}
with $\mathbf{D}_{\Omega}$ defined in \eqref{eqn: D_omega}.

\section{Proof of Theorem~\ref{theo: SIMO_general_N_special_corr}}\label{app: SIMO_MMSE_N_general_Theorem_proof}

In this case, \eqref{eqn: SIMO_1}--\eqref{eqn: SIMO_2} remain valid and, hence, \eqref{eqn: special_h_MMSE_D_I=0} holds with $\mathbf{A}=s\mathbf{I}_{N_R}$ and $\mathbf{D}_{\Omega}^{-\frac
{1}{2}}=\frac{1}{\sqrt{|s|^2+\sigma^2}}\mathbf{I}_{N_R}$. Let $P_{\textsf
R}=\mathcal{P}(\boldsymbol{\Lambda}_{\mathsf{R}}\boldsymbol{\Omega}_{\rm b}\boldsymbol{\Lambda}_{\mathsf{R}})$ and $P_{\textsf
I}=\mathcal{P}(\boldsymbol{\Lambda}_{\mathsf{I}}\boldsymbol{\Omega}_{\rm b}\boldsymbol{\Lambda}_{\mathsf{I}})$. Each off-diagonal element of
$\boldsymbol{\Lambda}_{\mathsf{R}}\boldsymbol{\Omega}_{\rm b}\boldsymbol{\Lambda}_{\mathsf{R}}$ (resp. $\boldsymbol{\Lambda}_{\mathsf{I}}\boldsymbol{\Omega}_{\rm b}\boldsymbol{\Lambda}_{\mathsf{I}}$) is affected by two different elements from $\mathbf{r}_{\mathsf{R}}$ (resp. $\mathbf{r}_{\mathsf{I}}$). Specifically, the $(i, k)$-th element of the standardized version of $\boldsymbol{\Lambda}_{\mathsf{R}}\boldsymbol{\Omega}_{\rm b}\boldsymbol{\Lambda}_{\mathsf{R}}$ is given by $\big(r_{\mathsf{R}, i}\sqrt{\frac{|s|^2\rho}{|s|^2+\sigma^2}}\big)\big(r_{\mathsf{R}, k}\sqrt{\frac{|s|^2\rho}{|s|^2+\sigma^2}}\big)$, for $i,k=1, \ldots, N_R$ and $i\neq k$.   Based on~\cite[p. 192]{Tong_MVN_book} and using \eqref{eqn: OrthantP_identity}, $P_{\textsf
{R}}$ is given by \eqref{eqn: MMSE_SIMO_same_rho_P_denom} and $P_{\mathsf{I}}$ is similarly obtained. Let $v_{\mathsf{R}, k}=\mathcal{P}\big(\big((\boldsymbol{\Lambda}_{\mathsf{R}}\mathbf{D}_{\mathsf{R}}\boldsymbol{\Lambda}_{\mathsf{R}})_{-k}\big)^{-1}\big)$ and  $v_{\mathsf{I}, k}=\mathcal{P}\big(\big((\boldsymbol{\Lambda}_{\mathsf{I}}\mathbf{D}_{\mathsf{R}}\boldsymbol{\Lambda}_{\mathsf{I}})_{-k}\big)^{-1}\big)$, for $k=1, \ldots, N_R$. By assumption, $\boldsymbol{\Sigma}$ is a standardized circulant matrix with all the off-diagonal elements equal to $\rho$. The inverse of $\boldsymbol{\Sigma}$ is also a circulant matrix given by~\cite{R_M_Gray}
\begin{align}
\boldsymbol{\Sigma}^{-1}=\frac{\left(1+(N_R-2)\rho\right) \boldsymbol{\Sigma}_{\rm inv,circ}}{1+(N_R-2)\rho-(N_R-1)\rho^2},\label{eqn: SIMO_arbitrary_N_pf_2}
\end{align}
where $\boldsymbol{\Sigma}_{\rm inv, circ}$ is a standardized circulant matrix with off-diagonal elements equal to $-\frac{\rho}{1+(N_R-2)\rho}$. Clearly, $\boldsymbol{\Omega}_{\rm b}$ and its inverse $\mathbf{D}_{\mathsf{R}}$ (cf. \eqref{eqn: SIMO_2}) are also circulant matrices. Let $\mathbf{D}_{\mathsf{R}, -k}$ denote the matrix obtained by deleting the $k$-th row and column of $\mathbf{D}_{\mathsf{R}}$, for $k=1, \ldots, N_R$. Using \eqref{eqn: SIMO_arbitrary_N_pf_2}, the $(N_R-1)\times (N_R-1)$ circulant matrix $(\mathbf{D}_{\mathsf{R}, -k})^{-1}$ is given by
\begin{align}
\left(\mathbf{D}_{\mathsf{R}, -k}\right)^{-1}=
\frac{\left((|s|^2+\sigma^2)^2-|s|^4\rho^2\right)\mathbf{D}_{\rm inv,circ}}{|s|^2+\sigma^2}, \label{eqn: SIMO_arbitrary_N_pf_3}
\end{align}
where $\mathbf{D}_{\rm inv,circ}$ is a standardized circulant matrix with off-diagonal elements equal to $\frac{\rho|s|^2}{|s|^2(1+\rho)+\sigma^2}$. 
 It is now clear that the $(i, l)$-th element of the standardized version of
\begin{align}
\big((\boldsymbol{\Lambda}_{\mathsf{R}}\mathbf{D}_{\mathsf{R}}\boldsymbol{\Lambda}_{\mathsf{R}})_{-k}\big)^{-1}
=\boldsymbol{\Lambda}_{\mathsf{R}, -k}\left(\mathbf{D}_{\mathsf{R}, -k}\right)^{-1}\boldsymbol{\Lambda}_{\mathsf{R}, -k}
\nonumber 
\end{align} 
is given by $\big(r_{\mathsf{R},i}\sqrt{\frac{|s|^2\rho}{|s|^2(1+\rho) +\sigma^2}}\big)\big(r_{\mathsf{R},l}\sqrt{\frac{|s|^2\rho}{|s|^2(1+\rho) +\sigma^2}}\big)$, for $i, l=1, \ldots, N_R$, $i\neq l$, $i\neq k$, and $l\neq k$. Invoke the results in~\cite{Tong_MVN_book} again to obtain $v_{\mathsf{R}, k}$ as in \eqref{eqn: MMSE_SIMO_same_rho_P_vec_num} and $v_{\mathsf{I}, k}$ is similarly obtained. Now that all the orthant probabilities in \eqref{eqn: special_h_MMSE_D_I=0} have been derived, Theorem~\ref{theo: SIMO_general_N_special_corr} readily follows. \hfill \IEEEQED

\section{Proof of Theorem~\ref{theo: theo_rank_one_SISO}} \label{app: proof_of_theo_SISO_rank_one}

Recall \eqref{eqn: SISO_DR_DI_rank_one} with $\mathbf{D}_{\mathsf{I}}=\mathbf{0}$, so that \eqref{eqn: special_h_MMSE_D_I=0} holds.  In this case, $\mathbf{A}$ reduces to $\mathbf{s}$  and  $\boldsymbol{\Sigma}$ to $1$. Clearly, $\mathbf{D}_{\Omega}=\Diag{\big(\big[s_k^2+\sigma^2\big]_{k=1}^\tau\big)}$. Let $\overline{P}_{\textsf R}=\mathcal{P}(\boldsymbol{\Lambda}_{\mathsf{R}}\boldsymbol{\Omega}_{\rm b}\boldsymbol{\Lambda}_{\mathsf{R}})$ and $\overline{P}_{\textsf I}=\mathcal{P}(\boldsymbol{\Lambda}_{\mathsf{I}}\boldsymbol{\Omega}_{\rm b}\boldsymbol{\Lambda}_{\mathsf{I}})$. Also, let $\overline{v}_{\mathsf{R}, k}=\mathcal{P}\big(\big((\boldsymbol{\Lambda}_{\mathsf{R}}\mathbf{D}_{\mathsf{R}}\boldsymbol{\Lambda}_{\mathsf{R}})_{-k}\big)^{-1}\big)$ and $\overline{v}_{\mathsf{I}, k}=\mathcal{P}\big(\big((\boldsymbol{\Lambda}_{\mathsf{I}}\mathbf{D}_{\mathsf{R}}\boldsymbol{\Lambda}_{\mathsf{I}})_{-k}\big)^{-1}\big)$, for $k=1, \ldots, \tau$. Based on \eqref{eqn: special_h_MMSE_D_I=0}, we have
\begin{align}
\hat{h}_{\rm MMSE} & =\frac{\mathbf{s}^\Trans}{2\sqrt{\pi}}\Diag{\left(\left[\frac{1}{\sqrt{s_k^2+\sigma^2}}\right]_{k=1}^\tau\right)} \nonumber
\\& \phantom{=} \ \cdot\left(\frac{\boldsymbol{\Lambda}_{\mathsf{R}}\big[\overline{v}_{\mathsf{R}, k}\big]_{k=1}^\tau}{\overline{P}_{\mathsf{R}}}+\frac{j\boldsymbol{\Lambda}_{\mathsf{I}}\big[\overline{v}_{\mathsf{I}, k}\big]_{k=1}^\tau}{\overline{P}_{\mathsf{I}}}\right). \label{new_eqn: proof_theo4_added}
\end{align}
Each off-diagonal element of
$\boldsymbol{\Lambda}_{\mathsf{R}}\boldsymbol{\Omega}_{\rm b}\boldsymbol{\Lambda}_{\mathsf{R}}$ (resp. $\boldsymbol{\Lambda}_{\mathsf{I}}\boldsymbol{\Omega}_{\rm b}\boldsymbol{\Lambda}_{\mathsf{I}}$) is affected by two different elements from $\mathbf{r}_{\mathsf{R}}$ (resp. $\mathbf{r}_{\mathsf{I}}$). Based on \eqref{eqn: SISO_DR_DI_rank_one}, the $(i, k)$-th element of the standardized version of $\boldsymbol{\Lambda}_{\mathsf{R}}\boldsymbol{\Omega}_{\rm b}\boldsymbol{\Lambda}_{\mathsf{R}}$ is given by $\big(\frac{r_{\mathsf{R}, i}s_i}{\sqrt{s_i^2+\sigma^2}}\big)\big(\frac{r_{\mathsf{R}, k}s_k}{\sqrt{s_k^2+\sigma^2}}\big)$, for $i,k=1, \ldots, N_R$ and $i\neq k$.
According to~\cite[p. 192]{Tong_MVN_book}, $\overline{P}_{\mathsf{R}}$ is given by \eqref{eqn: orthant_P_SISO_W_R} and $\overline{P}_{\mathsf{I}}$ is similarly obtained. On the other hand, from \eqref{eqn: SISO_DRK_inverse}, the $(i, l)$-th element of the standardized version of $\big((\boldsymbol{\Lambda}_{\mathsf{R}}\mathbf{D}_{\mathsf{R}}\boldsymbol{\Lambda}_{\mathsf{R}})_{-k}\big)^{-1}$ is given by $\big(\frac{r_{\mathsf{R},i}s_i}{\sqrt{s_k^2+s_i^2+\sigma^2}}\big)\big(\frac{r_{\mathsf{R},l}s_l}{\sqrt{s_k^2+s_l^2+\sigma^2}}\big)$, for $i, l=1, \ldots, N_R$, $i\neq l$, $i\neq k$, and $l\neq k$. Again, based on~\cite{Tong_MVN_book}, $\overline{v}_{\mathsf{R},k}$ is given by \eqref{eqn: orthantP_SISO_Wx_k} and $\overline{v}_{\mathsf{I},k}$ is similarly obtained. Finally, we insert the above orthant probabilities into \eqref{new_eqn: proof_theo4_added} and  \eqref{eqn: h_MMSE_SISO_multipilot} follows. \hfill \IEEEQED
\end{appendices}
\section*{Acknowledgments}
The authors would like to thank the anonymous reviewers and the associate editor for
their helpful comments on this paper. The first author wishes to 
thank Dr. P. Dharmawansa for  helpful suggestions and discussions regarding this work.

\bibliographystyle{IEEEtran}

\end{document}